%% file: ReferendumAnalysis.tex
\documentclass[aps,a4paper,10pt,onecolumn]{revtex4-1}

\usepackage{amsthm,amsmath}
\usepackage{braket}
\usepackage[utf8]{inputenc} 
\usepackage{subfigure}
\usepackage{url}
\usepackage{graphicx}

\newcommand{\brkt}[2][\big]{#1( #2 #1)}

\begin{document}


\title{Political Discussion and Leanings on Twitter: \\
the 2016 Italian Constitutional Referendum}


\author{Jacopo Bindi}
\thanks{These authors contributed equally}
\email{}
\affiliation{Department of Applied Science and Technology, Politecnico di Torino, Torino, Italy}

\author{Davide Colombi}
\thanks{These authors contributed equally}
\email{}
\affiliation{Sorbonne Universit\'es, UPMC Univ Paris 06, INSERM, Institut Pierre Louis d'Epid\'emiologie et de Sant\'e Publique (IPLESP), Paris, France}
\affiliation{Department of Physics and INFN, University of Turin, Torino, Italy}

\author{Flavio Iannelli}
\thanks{These authors contributed equally}
\email{iannelli.flavio@gmail.com}
\affiliation{Institut f\"ur Physik, Humboldt-Universit\"at zu Berlin, Berlin, Germany}

\author{Nicola Politi}
\thanks{These authors contributed equally}
\email{}
\affiliation{Dipartimento di Matematica, Universit\`a degli Studi di Torino, Torino, Italy}
\affiliation{Dipartimento di Scienze Matematiche (DISMA), Politecnico di Torino, Torino, Italy}

\author{Michele Sugarelli}
\thanks{These authors contributed equally}
\email{}
\affiliation{Dipartimento di Fisica, Universit\`a di Roma Sapienza, Roma, Italy}

\author{Raffaele Tavarone}
\thanks{These authors contributed equally}
\email{}
\affiliation{IIT, Genova, Italy}

\author{Enrico Ubaldi}
\thanks{These authors contributed equally}
\email{}
\affiliation{ISI Foundation, Torino, Italy}

\begin{abstract} 
The recent availability of large, high-resolution data sets of online human
activity allowed for the study and characterization of the mechanisms shaping
human interactions at an unprecedented level of accuracy.
To this end, many efforts have been put forward  to understand how people
share and retrieve information when forging their opinion about a certain
topic.  Specifically, the detection of the political leaning of a person
based on its online activity can support the forecasting of opinion trends
in a given population.
Here, we tackle this challenging task by combining complex networks theory and
machine learning techniques. In particular, starting from a collection of more
than 6 millions tweets, we characterize the structure and dynamics of the
Italian online political debate about the constitutional referendum held in
December 2016. We analyze the discussion pattern between different political
communities and characterize the network of contacts therein. Moreover, we set
up a procedure to infer the political leaning of Italian Twitter users, which allows us
to accurately reconstruct the overall opinion trend given by official polls
(Pearson's $r=0.88$) as well as to predict with good accuracy the final outcome
of the referendum. Our study provides a large-scale examination of the Italian
online political discussion through sentiment-analysis, thus setting a baseline
for future studies on online political debate modeling.
\end{abstract}

\maketitle

\section*{Introduction}

In the last years, online social media have reached a fundamental role in the
political discussion as they allow to easily spread a slogan or a political
campaign in a large population of users.
Every time we access a piece of information, share a content or comment on a
news we leave a permanent digital trace, which can be used to infer our opinion
regarding one particular event or
topic~\cite{conover2011alignement,conover2011political,ciulla2012beating}.
Among the variety of social media services, the micro-blogging platform Twitter
is probably the most commonly used for political debate and by political
leaders. This platform had a relatively recent diffusion in Italy, where it is
used by nearly the $10\%$ of the Italian adult population and where the former
prime minister Matteo Renzi was the first Italian politician to substantially found his
public communication on tweets (the 140 characters-long messages published on
Twitter). Also, Twitter stream of data is still available to the public through
its application programming interface (API), making it the best candidate to
study and describe the online political debates in  a country.

In fact, previous works (mostly focused on the USA) found that the political
discussion on Twitter can be a good proxy for the overall opinion of the
population, and that Twitter data can therefore be used to infer the effects of
a political event or to accurately predict the outcome of a
vote~\cite{bovet2016predicting}. Moreover, several studies have highlighted the
non-trivial dynamics of the opinions on Twitter and within political blogs,
showing that politically active web users tend to aggregate in homogeneous
communities, divided by political ideas, and avoiding discussion with the
counterpart~\cite{vespignani2009predicting,vespignani2012modelling,conover2011political,conover2011alignement}.

In this work, we analyze the political debate on Twitter regarding the
Italian \textit{Referendum Costituzionale} (constitutional referendum) held on
December the fourth, 2016.
To this end, we firstly collected vote-related tweets during the intense
political discussion that took place on social media during the three months
before the vote. We then analyzed these tweets employing machine learning and
network theory techniques. To the best of our knowledge, we are the first to
apply sentiment-analysis to the political discussion on Twitter in the Italian
scenario.

We aim at the detection of communities composed by interacting users and at
characterizing the within-community homogeneity in terms of users opinions,
investigating to which extent one community is influenced by users of other
communities (outsiders).
Additionally, we explore the possibility to use Twitter data as a reliable
opinion poll and we assess their predictive power of the final outcome of the
vote.

We found the Italian political discussion to be no different from the more
studied USA case: strongly polarized communities act as echo-chambers and
internally speak via retweets. On the other hand, the intercommunity discussion
is mainly based on mentions used to report and criticize adversaries' quotes. We
also found that the temporal network structure generated by such contacts does
react to major events happened during the political campaign. Examples include
debates between the two major parties leaders held on TV and political or
juridical events connected to the referendum.

The work is structured as follows: in Data Collection we describe the
procedure adopted to retrieve the tweets and in Tweets classification the
development of the classifier used to predict the leaning of each tweet in the
dataset.
From this vantage point, in User dynamical opinion we define a procedure
to assign a dynamical opinion to a given user. These dynamical opinions are
then used to characterize the time-varying network of contacts
(Static characterization of the tweets network)  and to identify the most influential
spreaders of the online political debate (Network dynamics: influential spreaders identification). We
further leverage on the reconstructed users opinion in
Comparison with official polls, where we compare our recreated opinion trend
to the empirical signal given by a large set of official polls, finding a very
good agreement between the two. Finally, we draw the conclusions of our study.

\section*{Methods and Results}
$\;$
\subsection*{Data Collection}\label{sec:dataColl}
Starting from the midnight of the 30th of August 2016, we collected all the tweets
in Italian containing one or more of the following strings: \emph{renzi,
iovotono, iovotosi, referendum, referendumcostituzionale, bastaunsi, \#NO, \#SI,
riformacostituzionale, m5s, pd, costituzione}. Some of these strings are the
keywords of the political campaign for the two main opposite formations
(\emph{iovotono, iovotosi, bastaunsi, \#NO, \#SI}) or the name of the two main
political parties, while others may be linked to the political debate on the
referendum. To filter out tweets written in languages other than Italian we
relied on the automatic twitter language detection.

The collected tweets incorporate information on the author, the time of
creation, the location (if available), the text content, the mentions
(\textit{i.e.}, users cited in the text by inserting \emph{@username}), the
links and urls inserted, the hashtags, and the tweet kind.
The latter specifies if the tweet is an original tweet (\textit{i.e.} a new
content generated by the user) or a retweet, meaning that the user reposted on
its account a tweet previously generated by another user. A retweet contains a
unique identifier of the original tweet being reposted, its original text and
author.
We did not reconstruct the follower network of the users (i.e., the network
where two users are connected if one of the two is following the other) as we
only use the single retweets/mentions as a proxy for the contact between
individuals in the network.
Data were recorded until the midnight of December the 4th 2016, right after the
end of the consultation and the publication of the first exit polls.
The resulting data set consists in $6\,894\,389$ tweets, authored by a total of
$N_{users}=266\,437$ users.

Note that we did not filtered for accounts possibly associated to bots,
\textit{i.e.} accounts not run by humans but programmed to automatically post
and share information. These are known to be part of the traffic
volume~\cite{ratkiewicz2011truthy,ferrara2016rise,suarez2016influence}, but here
we assume that their net effect on the contact network is negligible.

\subsection*{Tweets classification}
\label{sec:tweetClass}

\begin{figure}
    \centering
    {\includegraphics[scale=0.6]{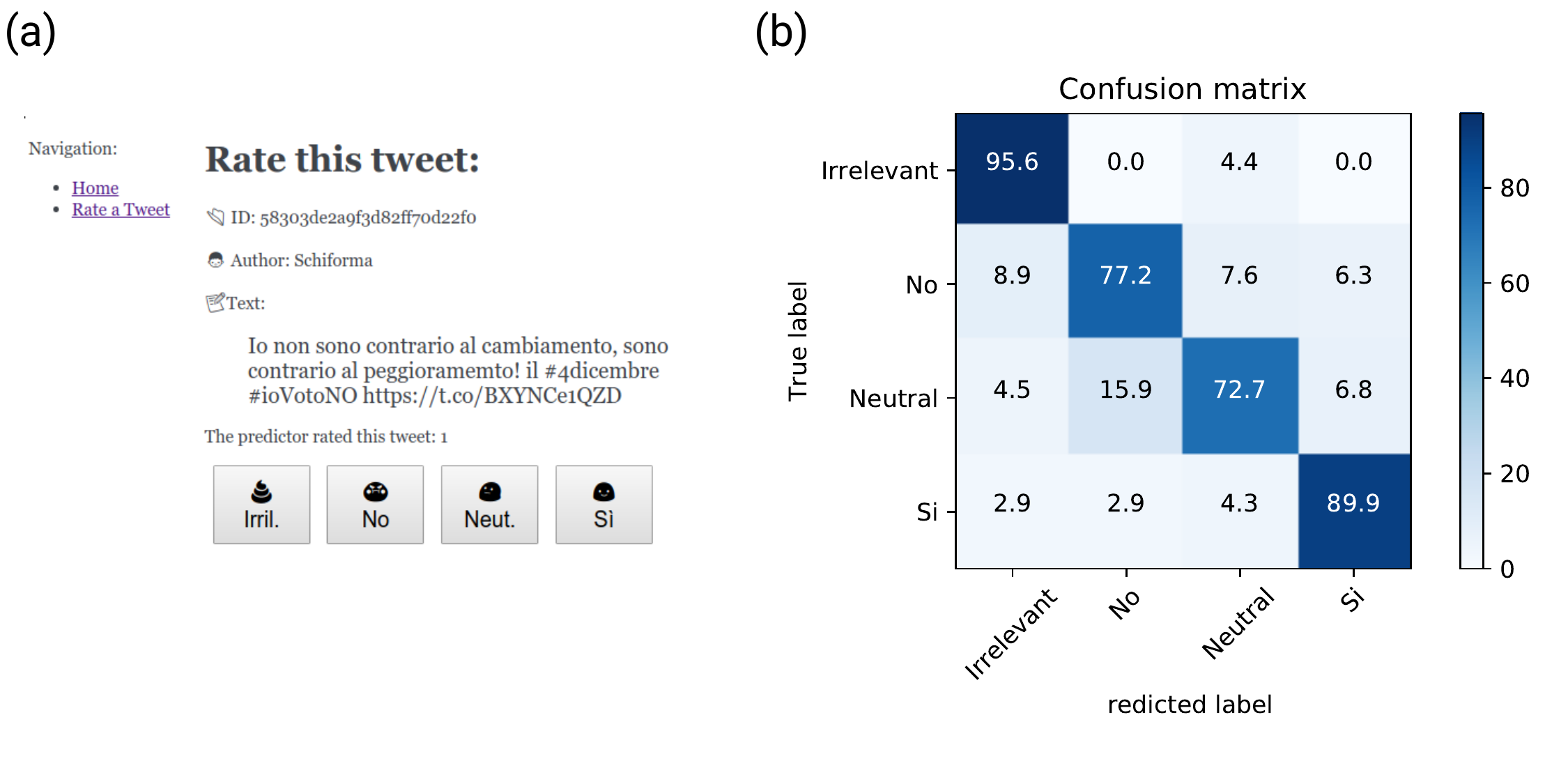}}
    \caption{
         (a) The web interface presented to the human voter containing the
         unique identifier of the tweet in the database, the author's nickname
         and the text of the tweet. If the tweet already features a preliminary
         classification, this is shown above the four buttons to classify the
         current tweet. Once the user inputs its preference, the system
         automatically presents a new tweet to be categorized.
         (b) The confusion matrix with percentage values for the random forest
         model with 21 estimators using the top 200 words and hashtags as
         features.
    }\label{fig:figure1}
\end{figure}

Given the political event under investigation, tweets need to be classified as
belonging to one of four following classes: irrelevant, pro-no, neutral, and
pro-yes.
Following \cite{salathe2011sentiment}, we rely on supervised machine learning
techniques to classify the tweets, for which we first needed an annotated data
set to train a classifier. Manual classification was made by developing a simple
web interface (shown in \figurename~\ref{fig:figure1} panel a) that prompts the user to
classify each tweet into one of the four given categories. The information
displayed to the user are: the text of the tweet, the contained hashtags, the
author name, and all the urls included in the tweet.

Because of the sometimes ambiguous inclination of a tweet and the subjective
perception of a tweet being more prone to the pro-yes or pro-no classes, we
strengthened the manual classification of tweets by considering a tweet to be
classified when \emph{i}) it received at least 3 votes and \emph{ii}) at least
two thirds of them are in agreement.
At the end of this procedure we ended up with a fairly balanced training set:
out of the 1150 labeled tweets, 306 have been voted as irrelevant, 375 pro-no,
203 as neutral and the remaining 266 as pro-yes.

\subsubsection*{Sentiment-analysis}

While previous works on the Italian scenario limited their attention to the
volume of tweets containing fixed hashtags as a proxy for the overall volume of
political affiliation~\cite{eom2015twitter,caldarelli2014multi}, in this work we
apply for the first time sentiment-analysis to Twitter data in the Italian
political discussion. Although the former approach is certainly pioneering, it
has several counter-effects. For example, a  pro-yes tweet containing the
official pro-no hashtag would straightforwardly increase the pro-no volume. Our
method, detailed in the following, overcomes such difficulties.

The features of our model are the low-case words extracted from the text of the
tweets where we removed punctuation and we substituted all the posted urls with
their principal domain name (\textit{e.g.}, a link pointing to
\url{http://www.newspaper.it/politics/news_about_referendum} is replaced by
\url{newspaper.it}). Smileys are left as UTF-8 characters and we leave hashtags
as separate entities, \textit{i.e}. we do not treat them as simple words. This
last choice can be justified \emph{a-posteriori} by observing that the
classifiers accuracy drops significantly when projecting hashtags to words by
removing their leading pond \#.
We further process the features by stemming (using the
\texttt{WordNetLemmatizer} from the nltk module of python~\cite{nltkPython}) and
removing the Italian stop-words from the corpus (using the \texttt{stopwords}
collection from \texttt{nltk.corpus}). The resulting dictionary is composed by
535 different hashtags and 3773 words.

The manually annotated data set is used to train and validate three different
classifiers: logistic regression, naive Bayes, and random
forest~\cite{salathe2011sentiment,mostafa2012learning}.
Results are then compared to choose the best-performing model.  The performances
of each model are evaluated using a 4-fold cross validation scheme: a fourth of
the labeled tweets is randomly selected for validation and the remaining  three
fourths for training, and the procedure is repeated four times. The accuracy of
a model is defined as the 4-fold average of the percentage of tweets in the
validation set whose automatic classification correctly matches the manual one.
The three classifiers give relatively similar performances: 80$\%$ accuracy for
the random forest model, 82$\%$ accuracy for logistic regression and 79$\%$
accuracy for the naive Bayes classifier.

To gain better performances we further improved the random forest classifier by
leveraging on the weights of the fitted model that naturally measure the
importance of each word.
The procedure is implemented as follows:
\begin{itemize}
    \item we trained a random forest classifier using the 3773 words in our
        dictionary (hashtags excluded) as features. The model fitted in this way
        has relatively low predictive power, with accuracy $\sim 50\%$;
    \item we selected the top 200 words ranked by feature importance (the top 10
        most predictive words are \emph{referendum, renzi, no, s\`i, riforma,
        m5s, salvini, paese, grillo});
    \item finally, we trained a random forest model using all the hashtags and
        the top 200 words as features, obtaining 86$\%$ accuracy using 21
        estimators and an impurity split equal to $3.1\cdot10^{-6}$.
\end{itemize}

The confusion matrix of the model on the annotated dataset is shown in
\figurename~\ref{fig:figure1} panel b.
The model was finally used to classify all the remaining tweets, resulting in
$\approx 2.6M$ irrelevant, $\approx 1.7M$ pro-no, $\approx 1.1M$ neutral and
$\approx 0.5M$ pro-yes tweets.

\subsection*{User dynamical opinion}
\label{sec:opinionDef}

In the current section, we show and discuss the procedure used to define the
opinion of each user in each day of the data collection, based on the users'
classified tweets.
We denote by $i\in \{1,\dots, N_{users}\}$ the index identifying a given user,
by $T=97$ the length of the data set expressed in days, and by $t \in
\{1,\dots, T\}$ a given day ($t=1$ corresponds to August 31st and $t=97$ to
December the 4th, the date of the referendum). Then, we denote by
$\mathrm{(Yes)}_{i,t}$ the number of pro-yes tweets posted at time $t$ by author
$i$, while the quantity  $\mathrm{(No)}_{i,t}$ is defined analogously. Note that
tweets classified as irrelevant are considered equivalent to those classified as
neutral in what follows.

We define the daily activity as the opinion that the user expresses the most in
its daily tweets, namely
\begin{equation}\label{DefOp:eq:daily_activity}
    a_{i,t} = 
    \begin{cases}
        +1 & \quad\mathrm{if}~ \mathrm{(Yes)}_{i,t} > \mathrm{(No)}_{i,t} \\
        -1 & \quad\mathrm{if}~ \mathrm{(Yes)}_{i,t} < \mathrm{(No)}_{i,t} \\
        0  & \quad\mathrm{if}~ \mathrm{(Yes)}_{i,t}=\mathrm{(No)}_{i,t} \, .
    \end{cases}
\end{equation}
Note that the third condition also implies that if user $i$ either posts only
neutral/irrelevant tweets or does not tweet at all, then its daily activity is
assumed to be neutral.

The daily activity defined in \eqref{DefOp:eq:daily_activity} is not suitable to
represent a user's opinion, because it only captures what the user tweeted on a
given day.
Indeed, it is reasonable to assume that a person maintains her/his opinion even
though she/he does not declare it on Twitter every single day. For such a
reason, we define an opinion function that maps the daily activity of user $i$
to $o_{i,t}$, the actual opinion of user $i$ at time $t$. To include a memory
effect, the opinion of user $i$ is the projected weighted sum of its daily
activity up to time $t$
\begin{equation}\label{DefOp:eq:opinion_definition}
    o_{i,t} = p\brkt[\Bigg]{\sum_{s=0}^{t} w_{s,t} a_{i,s}},
\end{equation}
where the projection function $p$ is discussed below and the $w_{s,t}$ are
exponentially decaying weights given by
\begin{equation*}
    w_{s,t} = \frac{ e^{-(t-s)/K} }{\sum_{s=0}^{t} e^{-(t-s)/K} }.
\end{equation*}
Here, parameter $K$ gives an approximation of the typical number of days before
$t$ that are taken into account when determining the opinion of the user. The
intuition is that the more recent a tweet is, the more it affects the opinion
value of the user.

The function $p$ projects the argument within the parenthesis in
\eqref{DefOp:eq:opinion_definition} onto the discrete opinion set $\{-1,0,+1\}$,
and a natural choice for such projection is therefore the sign function.
However, an undesirable drawback of setting $p(x)=\mathrm{sign}(x)$ is that even
a mildly polarized user who tweeted in favor of, say, the yes vote on the very
first day and then always posted neutral tweets for the following three months,
would maintain a $+1$ opinion forever. Indeed, even though weights $w_{0,t}$
rapidly approach zero as $t$ becomes large, a sign $p$ function would still
project the resulting small (but positive) sum on a pro-yes opinion.
Then, we consider the projection to be a sign-like function with a
$\epsilon$-widened preimage of zero, \textit{i.e.}
\begin{equation*}
    p_\epsilon(x) =
        \begin{cases}
            -1  &   \mathrm{if}~x\in[-1,\epsilon]  \\
            0   &   \mathrm{if}~x\in(-\epsilon,\epsilon)    \\
            +1  &   \mathrm{if}~x\in[\epsilon,1] \, .
        \end{cases}
\end{equation*}

Overall, two parameters are involved in the definition of the users opinion
\eqref{DefOp:eq:opinion_definition}: the decaying time of the weights $K$ and
$\epsilon$, the widening parameter of the projection function.
Since we set the parameter $K$ to be the typical number of days that a user
retains its opinion, we infer it from the data as follows. We fix $K$ to be the
average number of days between two coherent tweets authored by the same user
which are not separated by a differently classified tweet. Considering that on
average users tweet coherently every $5$ days and $7$ hours (without showing any
different daily activity in between), we deduce that the personal opinion is
preserved for $K=5.29$ days on average.
On the other hand, the role of the parameter $\epsilon$ is less self-evident.
We recall that such parameter is introduced to allow users to return
on a neutral position after a period of inactivity, but it actually has a more complex effect on the time evolution of the users opinion $o_{i,t}$ under several aspects. 
Here, we only report that the parameter is set to $\epsilon=0.075$ 
and we refer the reader to the Supplementary Material for a detailed discussion of its assessment. 

\begin{figure}
    \centering
    {\includegraphics[scale=0.50]{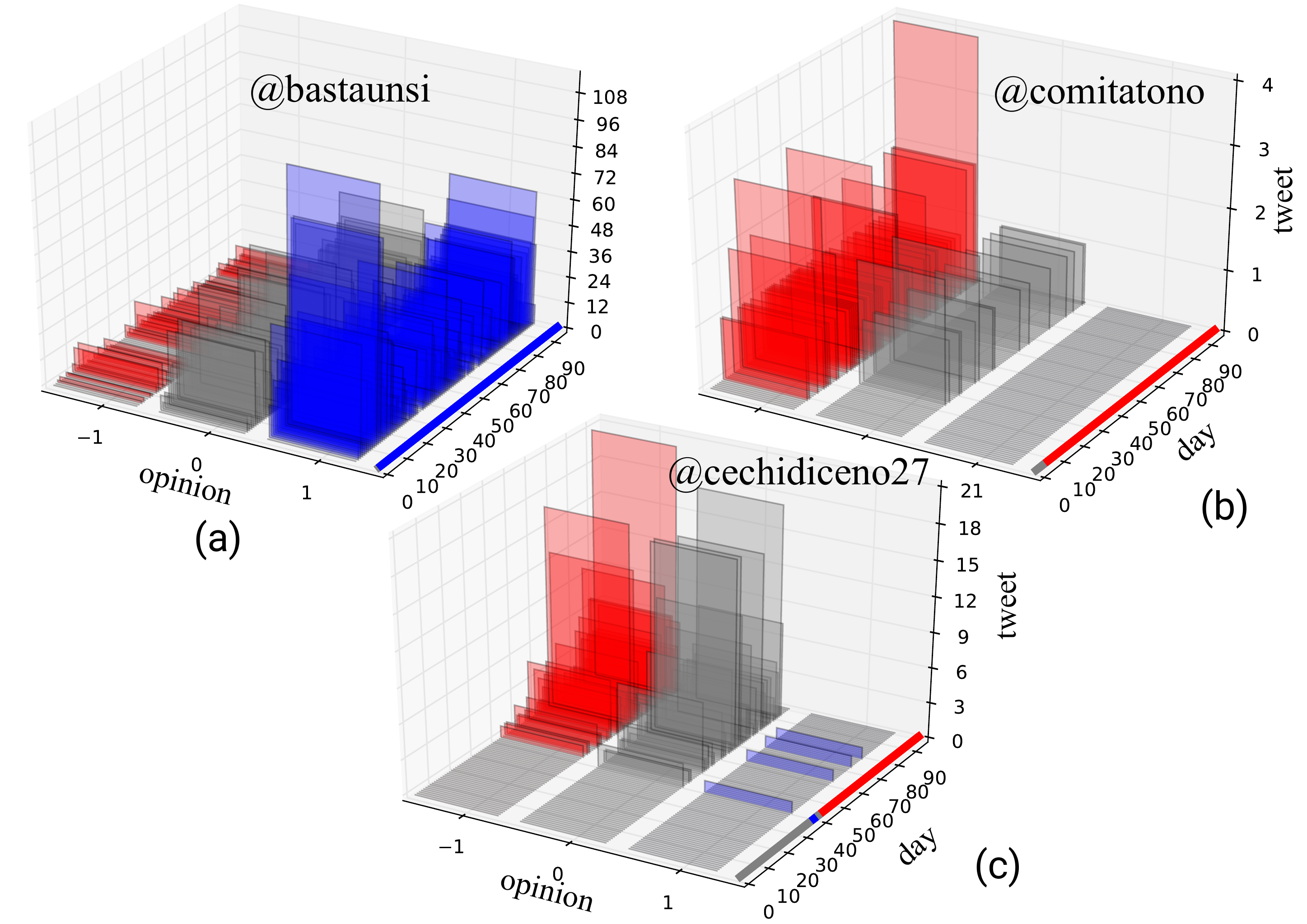}}
    \caption{
        \label{fig:figure2}
        Histogram representation of the number of tweets authored by different users,
        subdivided in number of tweets classified as pro-no ($-1$, red), neutral or
        irrelevant ($0$, gray), and pro-yes ($+1$, blue). The coloured line on the right of each panel
        shows the resulting opinion of the user as defined in
        \eqref{DefOp:eq:opinion_definition} with $K=5.29$ days and $\epsilon=0.075$.
        Panel (a) refers to the official pro-yes committee's account (@bastaunsi), panel
        (b) to the official pro-no committee's account (@comitatono), and panel (c) to
        user @cechidiceno27 that exhibits an opinion switch from a temporary pro-yes to
        a sustained pro-no leaning.
    }
\end{figure}

To illustrate the resulting opinion time course, we plotted the daily activity histograms
of a few users and the resulting opinion time course in
\figurename~\ref{fig:figure2}. A detailed description of the figure
content is given in the Supplementary Material.

Now that we defined the procedure to evaluate the opinion of each single user at
a given time $t$ we can investigate how the topology of the network of contacts
among Twitter users is shaped by the leaning of the users themselves.

\subsection*{Static characterization of the tweets network}
\label{sec:tweetsNetwork}

Starting from our dataset of annotated tweets we build two different networks:
\emph{i}) the retweet network (RN), where a directed edge is drawn from user B
to A when user A retweets user B, and \emph{ii}) the mention network (MN), where
we put a directed edge from user A to user B when A mentions B in a tweet. The
two conventions on the edge direction are adopted so as to reproduce in the
synthetic system the information flow found in the empirical social network.

The topology of this networks reflects the structure of the political debate
within Twitter, and it is studied on two levels. First, we measure the size and
track the evolution of the network connected components. We then detect the
communities present in the system aiming at the analysis of their political
polarization. Finally, we characterize the preferred mean of communication
between communities with diverse average opinion.

\begin{figure}
    \centering
    {\includegraphics[scale=0.3]{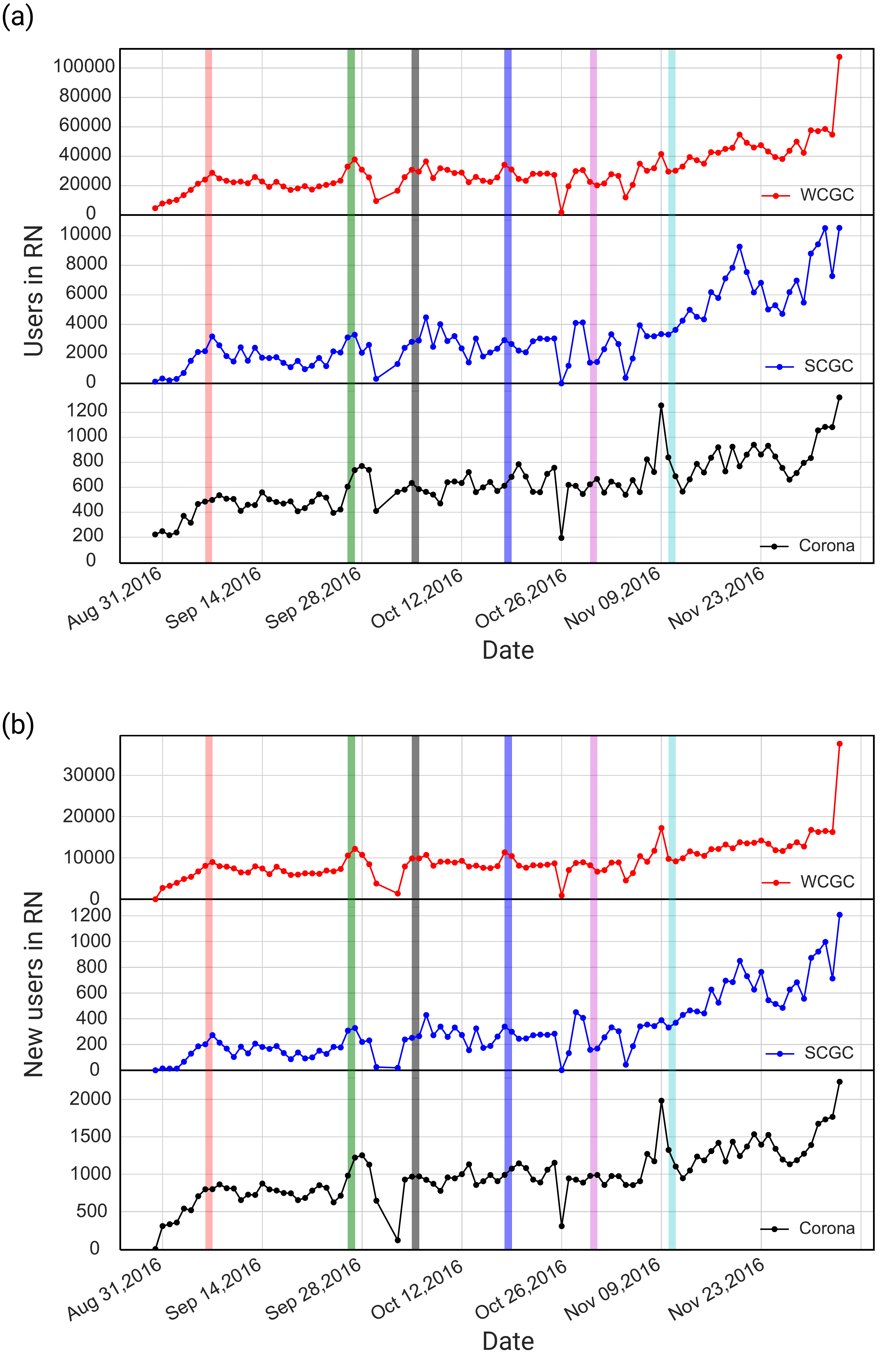}}
    \caption{
        \label{fig:figure3}
        Connectivity of the RN. (a) Total number of users versus time in the
        WCGC, the SCGC and the corona. (b) Number of new users (\textit{i.e}.,
        users seen for the first time in the system) entering as part of the
        WCGC, the SCGC and the corona. The WCGC is an order of magnitude bigger
        than the SCGC, and approximately represents the entire network.
        The vertical bands represent some events that had a significant impact
        for the referendum debate:  (red) the mayor of Rome, who previously
        endorsed the No, is involved in legal issues; (green) the Italian
        government fixes the referendum day; (black) the regional court of the
        region Lazio receive an appeal to invalidate part of the Referendum
        question formulation; (purple) the public debate about the referendum
        reaches the first pages of the main Italian newspapers; (pink)
        television debate with the Italian prime minister; (cyan) an important
        national meeting, Leopolda, organized by the Government party, is held
        in Florence.
    }
\end{figure}

\subsubsection*{Political communication network}

To characterize the connectivity properties of the whole network, we study the
weakly connected giant component (WCGC), the strongly connected giant component
(SCGC) and the corona~\cite{bovet2016predicting}.
The SCGC~\cite{dorogovtsev2001giant} is the core of a directed graph, and it is
constituted of every pair of vertices that are connected in both directions.
Thus, from one vertex in the SCGC one can approach any other vertex in the SCGC
by moving either along or against the edge directions (interactions loop). The
WCGC is a connected component formed by users that do not necessarily have
reciprocal interaction with each other, whereas the corona is the rest of the
network, \textit{i.e.} it is formed by the smaller components of the network
that do not belong to neither one of the giant components.

\begin{figure}
    \centering
    {\includegraphics[scale=0.3]{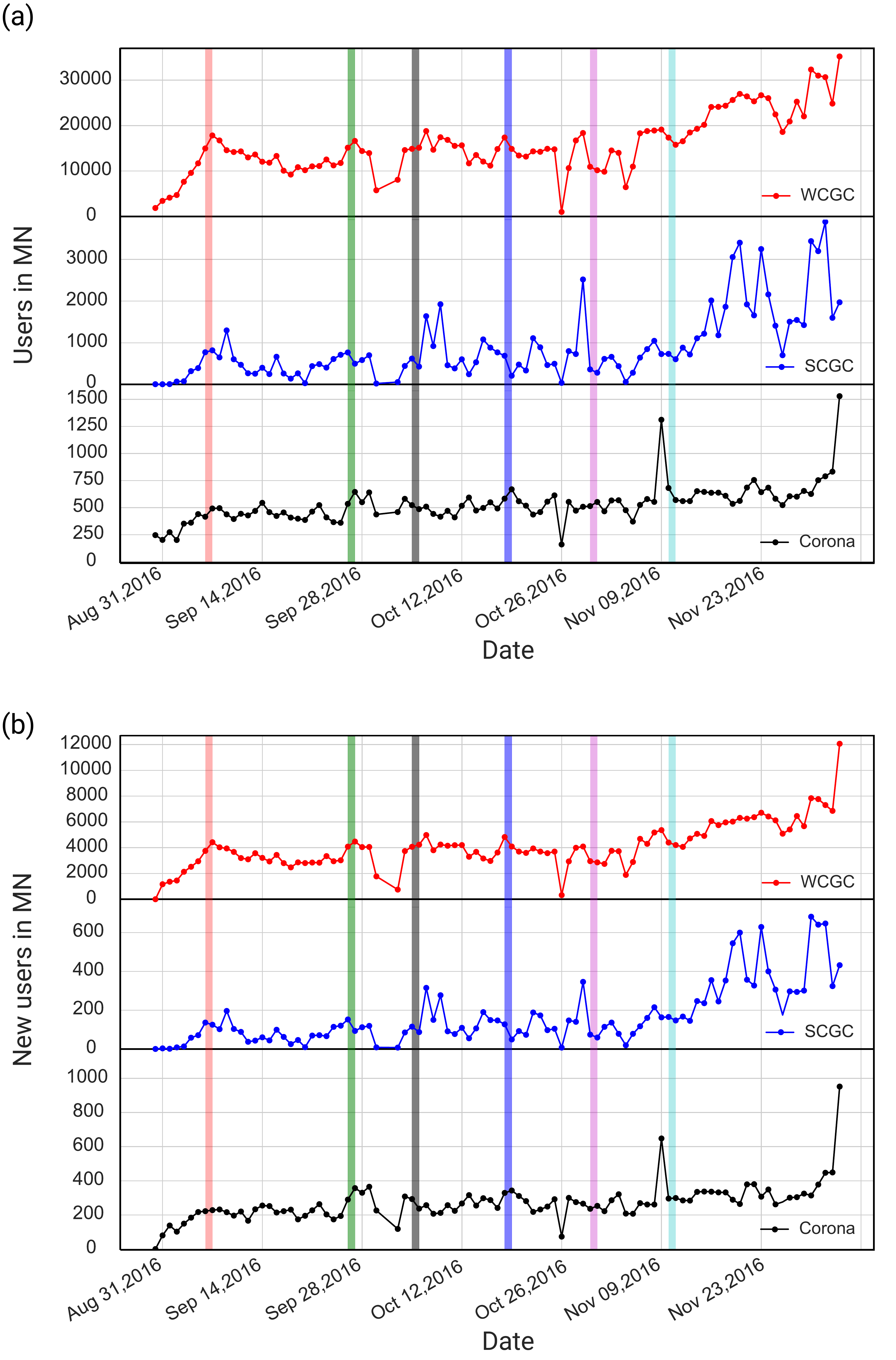}}
    \caption{
        \label{fig:figure4}
        Connectivity of the MN. (a) Total number of users belonging to the WCGC,
        the SCGC and the corona as a function of time. (b) Number of new users
        entering the system in the WCGC, in the SCGC and the corona. The WCGC is
        an order of magnitude bigger than the SCGC, and it approximately
        represents the whole network.
        The vertical bands represent some significant events for the referendum
        debate, as described in the caption of \figurename~\ref{fig:figure3}.
    }
\end{figure}

In the RN, the WCGC size ranges from $10^3$ to $4\times10^4$ users, while the
SCGC from $40$ to $1.6\times10^3$, as shown in \figurename~\ref{fig:figure3}. In the
MN  (see \figurename~\ref{fig:figure4}), the WCGC have a size between $700$ and
$1.7\times10^4$, and the SCGC between $5$ and $10^3$. Thus, most of the entire
set of nodes in both graphs belong to the WCGC (the coronas have an irrelevant
size).

To understand how new users join the conversation, we show the number of users
that appear for the first time in our network at each time step in
\figurename~\ref{fig:figure3} panel b and \figurename~\ref{fig:figure4} panel b.
As expected, most of the new users arrive in the WCGC and the corona. Indeed,
new users are more likely to be less active in the discussion. As the referendum
day approaches, both the SCGC size and the number of new users grow.
We also observe different spikes in the number of both active and new users.
These spikes correspond to particular dates, when important events stimulated
the debate on the referendum (vertical bands in \figurename~\ref{fig:figure3} and
\figurename~\ref{fig:figure4}) triggering a clear response of the network. We elaborate further on this topic in the following sections.

\subsubsection*{Relation between network's topology and users opinion}

\begin{figure}
    \centering
    {\includegraphics[scale=0.30]{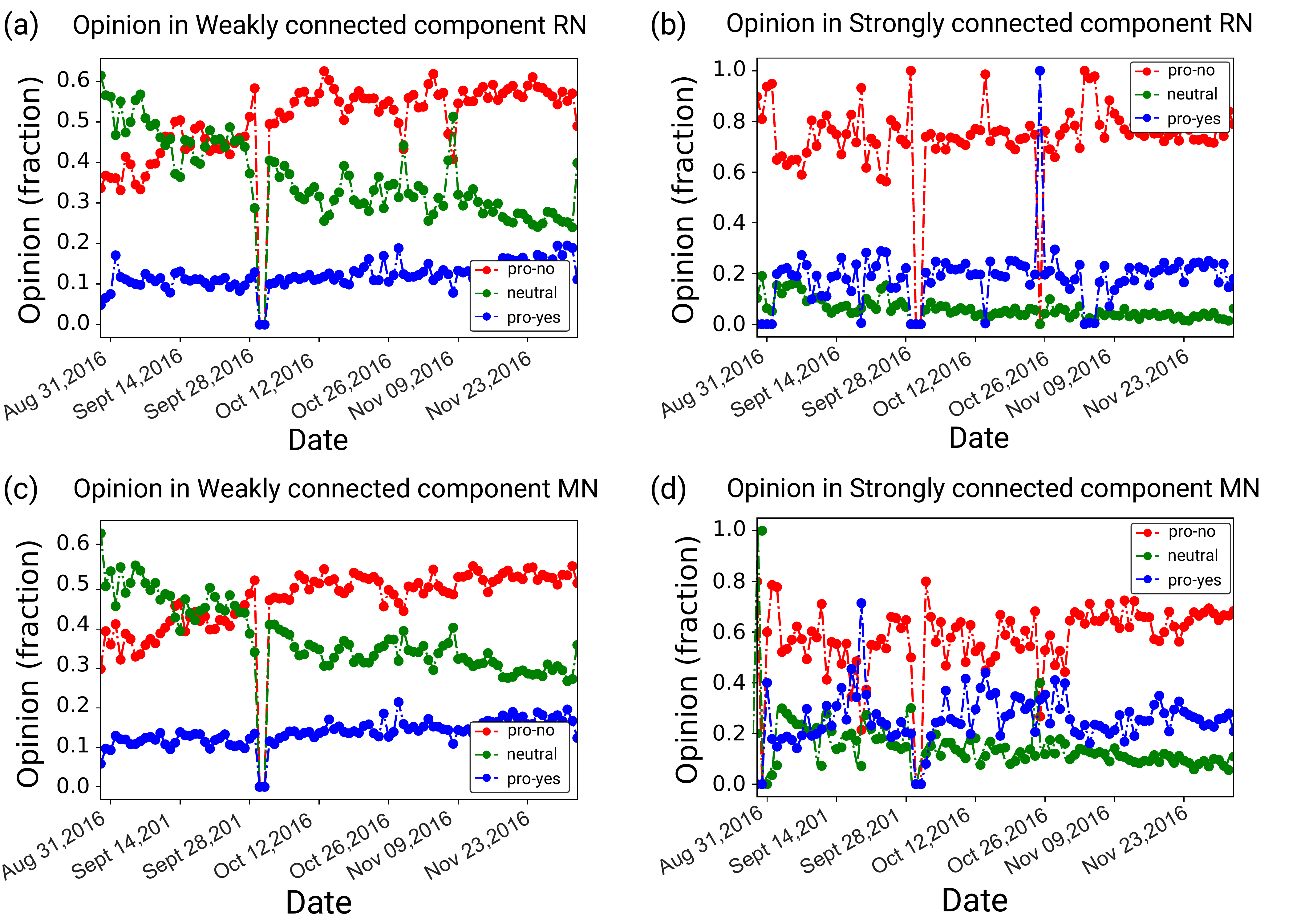}}
    \caption{
        \label{fig:figure5}
        Top: average opinion in (a) the weakly connected  and in (b) the
        strongly giant component in the RN. Bottom: average opinion in the  (c)
        weakly connected component  and in (d) the strongly giant component in
        the MN. The green dots are for the neutral opinion, red for the pro-no
        opinion, blue for the pro-yes opinion. Lines are guide to the eyes.
    }
\end{figure}

Given the network components found in the previous section we can now
investigate how the users' political opinion defined in
User dynamical opinion is distributed within them.
We find that the users with a neutral opinion are an important fraction of the
weakly connected component for both the RN and MN (Figure~\ref{fig:figure5} panels a
and c). As one can reasonably expect, the number of neutral
users decreases in time, as users opinions become more polarized while
approaching the voting day.
Additionally, polarized users tend to be more involved in interaction loops as it can be seen in Figure~\ref{fig:figure5} panels b and d.

\subsubsection*{Community influencers and coherence}

In order to gain further understanding of the political structure of the
networks we implemented a community detection algorithm, applying the Louvain
method \cite{blondel2008fast}.
The Louvain algorithm is a greedy optimization method that finds the optimal
division of a network into communities by iterative optimization of the network
modularity, a measure of the density of links between nodes within a community
with respect to the density of links outside it.
The method returns a set of subgraphs more densely connected to one another than
to other nodes, along with a hierarchy of communities at different scales.

\begin{figure}
    \centering
    {\includegraphics[scale=0.6]{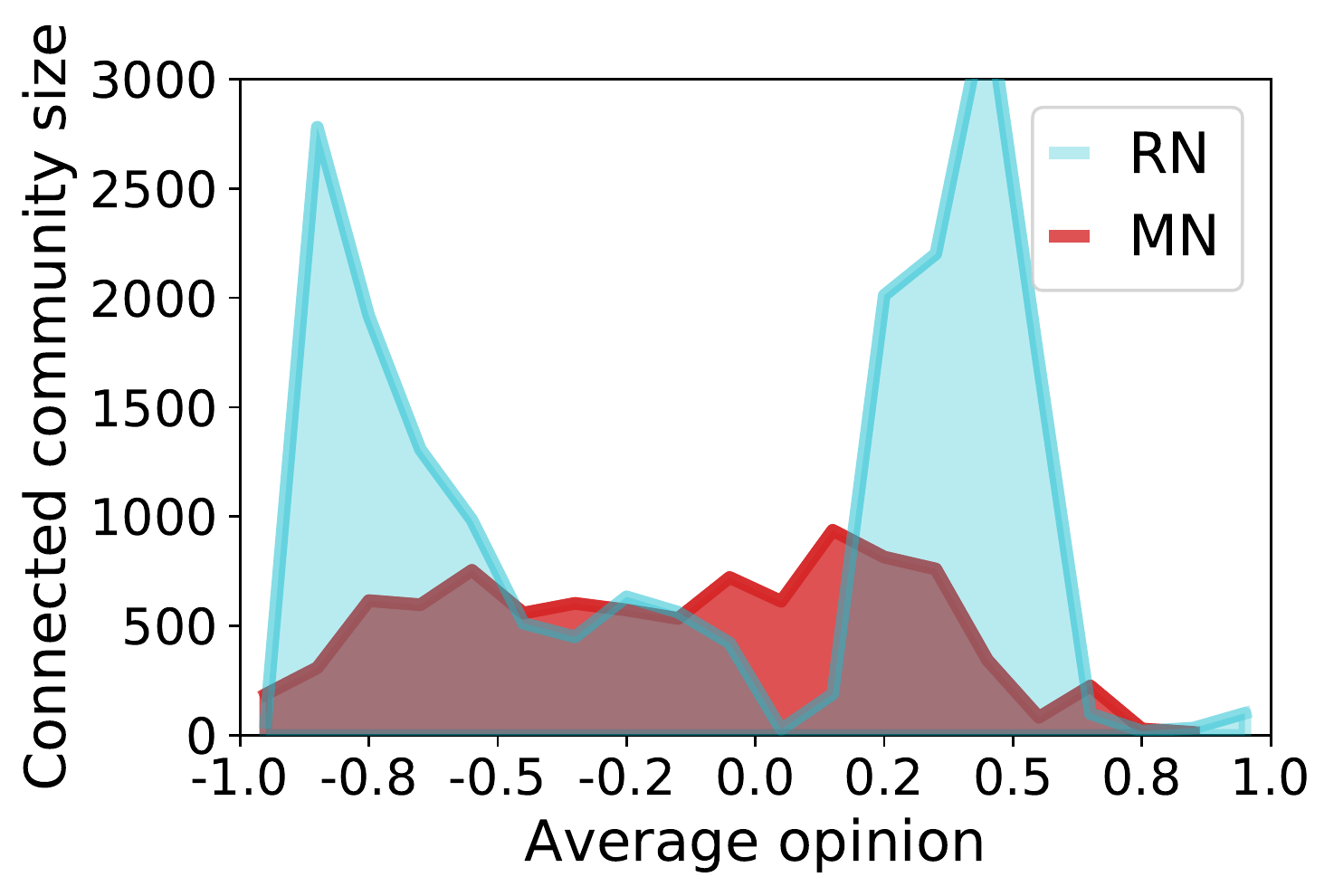}}
    \caption{
        \label{fig:figure6}
        The size of communities plotted as a function of 
        $\overline{O}_C$, the average opinion of users belonging to community $C$, for the RN (cyan
        area) and the MN (red area). In both cases, we show the
        $95$-th percentile of the community size distribution found for communities with a given average opinion.
    }
\end{figure}

Studying the size of the detected communities with respect to the average
opinion inside the community gives an insight on the network political
composition.
We measure the average opinion $\overline{O}_C$ inside a community $C$ as the mean
value of the opinion $o_{i,t}$ among the users belonging to the $C$ community
and over the full observation period. Thus:
\begin{equation}
    \overline{O}_C = \frac{1}{|C|T} \sum_{i \in C} \sum_{t=0}^{T}{o_{i,t}},
\end{equation}
where $|C|$ is the number of nodes in community $C$.
The result of such analysis are shown in \figurename~\ref{fig:figure6}.
In the RN the larger communities are those with a strongly polarized average
opinion, whereas we find the communities in the MN to have a weaker opinion
polarization, as the larger MN communities feature $\overline{O}_C \simeq 0$.
Interestingly, we find the opinion polarization to be stronger for the pro-no
communities of the RN. These communities are generally larger and more polarized
(\textit{i.e.} with more negative opinion) with respect to the ones supporting
the pro-yes faction. Moreover, we observe also in the MN case a slight shift of
the communities opinion toward the negative pole. This shift is reasonably due
to the overall predominance of the pro-no side which were against the governing
party (which was, on the contrary, promoting the pro-yes side).

\subsection*{Network dynamics: influential spreaders identification}
\label{sec:rumorSpreading}

After the analysis of the networks topology, it is worth considering how the
edges arrangement and their activation patterns shape the evolution of a rumor
spreading process unfolding on the network fabric. Specifically, in this
section we quantitatively characterize the dynamics of a rumor spreading (RS) model
on top of the mention and retweet networks.
Since in a reaction-contact spreading model on a directed graphs each node
(user) may spread and receive rumors only within the strongly connected
component, we restrict our analysis to that subset only.

To perform the rumor spreading dynamics we define the unweighted temporal
network (for both MN and RN), using the adjacency matrix $\mathbf{A}^{(t)}$
($t=1,\dots,T$) where each non-zero entry is set equal to $1$.
In Table~\ref{tab:widgets} we show the relevant network properties for the
undirected aggregated versions of the full and the SCGC graphs in both the MN
and the RN.
The high value of the second moment of the degree distribution, $\braket{k^2}$,
observed in the two full graphs (in particular in the MN) highlights the rapid
increase of the topological fluctuations, which corresponds to a high
heterogeneity H \cite{chen2012identifying}. This has important consequences on
the dynamical processes as it implies a rapid decrease of the epidemic threshold
in a corresponding standard infection process \cite{barrat2008dynamical}.

\begin{table}
    \centering
    \begin{tabular}{l|r|r|r|r|r|r|r}
        & Nodes & Edges & Density & D & C & $\braket{k}$ & H\\ \hline
        \hline
        Full MN & 80030 & 629061 & $ 9.821 \ 10^{-5}$ & $\infty$ & 0.22 & 15.72 & 70.32 \\

        SCGC MN & 15294 & 353867 & $ 1.513 \ 10^{-3}$ & 6 & 0.34 & 46.27 & 11.55  \\

        Full RN & 179680 & 1543963 & $ 4.782 \ 10^{-5}$ & $\infty$ & 0.11 & 17.19 & 47.19 \\

        SCGC RN & 27437 & 917959 & $ 1.219 \ 10^{-3}$ & 7 & 0.28 & 66.91 & 8.56
    \end{tabular}
    \caption{
        \label{tab:widgets}
        Statistical properties of the full and  SCGC aggregated MN and RN: the
        diameter D, the clustering C, the average degree $\braket{k}$ and the
        degree heterogeneity H defined as  $H=\braket{k^2}/\braket{k}^2$
        \cite{chen2012identifying}. All quantities  are obtained using the
        undirected counterparts of both time-aggregated graphs.
    }
\end{table}

\begin{figure}
    \centering
    {\includegraphics[scale=0.40]{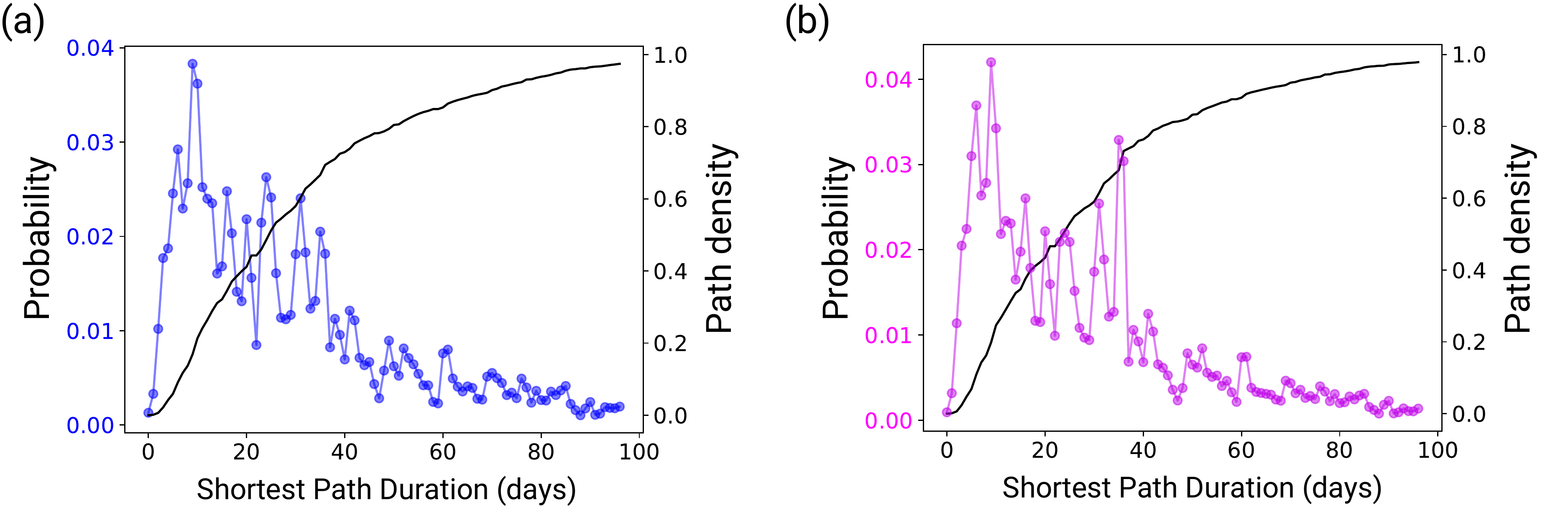}}
    \caption{
        \label{fig:figure7}
        Distribution of the shortest path duration and the density $\rho(\mathbf{\mathcal{A}}^T)$ of the accessibility graph (black) for (a)
        the SCGC of the MN (blue) and (b) the SCGC of the RN (violet). Causal fidelity values are $c = 0.973$
        and $c = 0.979$ for the MN and the RN, respectively.
    }
\end{figure}

Following \cite{lentz2013unfolding}, we compute the causal fidelity $c\in[0,1]$
of the SCGC graphs (see Supplementary Material).
We find that $c = 0.973$ and $c = 0.979$ for the MN and RN, respectively. These
values suggest that the temporal causality-driven effects in both SCGC networks
are practically negligible. Thus, it is reasonable to characterize the rumor
dynamics only considering the time-aggregated representations of the SCGC.
Figure~\ref{fig:figure7} panels a and b show the probability to find a
path of length $t$ between two randomly chosen nodes, and the corresponding
density of the accessibility graph.
In spite of the networks directness, the density of the accessibility graph
(black line) shows that after only half of the observation period (about 50
days) more than 80\% of the network is causally connected.
Therefore, the aggregated network representations give a good approximation of
the temporal one, in accordance with the high causal fidelity.
The shortest path length is broadly distributed for both the MN and RN, while its most frequent value is in both cases attained at a nine days long path. This means that the typical spreading time
scales are of the order of 10 days.

For the RS dynamics we consider the ignorant-spreader-stifler (ISS) model, with
a single initial spreader per simulation.
We focus on the Maki and Thomson ISS model~\cite{maki1973mathematical}, where
the rumor spreads via directed contacts. The directness feature of both MN and
RN plays indeed a fundamental role in the corresponding evolution of the ISS
dynamics as many users will never be able to spread a rumor that can reach a
substantial portion of the network. The reaction dynamics for adjacent users is
defined by the following spreading model \cite{barrat2008dynamical}
\begin{equation}
    \begin{cases}
        X + Y \xrightarrow{\beta } 2Y \\
        Y + Z \xrightarrow{\mu } 2Z \\
        Y + Y \xrightarrow{\mu } Z + Y
    \end{cases}
    \label{iss}
\end{equation}
where $X^{(t)}$, $Y^{(t)}$ and $Z^{(t)}$ are the ignorant, spreader and stifler
compartments, respectively, and the total population is $N = X^{(t)} + Y^{(t)} +
Z^{(t)}$.
In the Twitter case, spreaders correspond to users that have an information
(such as a specific news about referendum) and if one spreader meet an ignorant
user then the latter begins to spread this rumor as well. On the other hand,
stiflers are users that lose interest in the news and persuade spreaders to stop
propagating the rumor. Finally, the presence of two spreaders together could
bring one of them to became a stifler.
The two free parameters are the spreading (infection) and the stifler
(recovery) rates $\beta$ and $\mu$. The main difference to the standard
susceptible-infected-recovered (SIR)~\cite{anderson1992infectious} model is the
absence of spontaneous recovery: the transition to the stifler class may happen
through a contact with another stifler or via a contact between two spreaders.

The problem of identifying the most influential nodes in a network,
\textit{i.e.} the so-called top spreaders, is a central issue in network
science. Many works have addressed the question starting with the seminal paper
of Kitsak \emph{et al.} \cite{kitsak2010identification}, where the authors
showed how the k-shell decomposition of the graph can give a much more elaborate
description on how central a node is with respect to local measures.
The lack of a satisfactory understanding of the problem comes from the high
heterogeneity of the role played by the individual nodes in a complex network.
Centrality measures are then used to quantitatively derive the importance of
individual nodes \cite{wasserman1994social}.

Here we apply standard heuristic centrality metrics, namely the out-degree
$k^{out}$, betweenness $b$ \cite{freeman1977set}, closeness $cl$, eigenvector
centrality $e$ \cite{bonacich1972factoring}, k-core index $k_c$
\cite{seidman1983internal} and PageRank centrality $x$ \cite{page1999pagerank}.
Other non-heuristics centrality measures (as the non-backtracking centrality
\cite{martin2014localization}, which is supposed to match exactly the spreading
capacity at criticality in the standard SIR model)  cannot be used on directed
networks. Because of the highly directed nature of both RN and MN we do not
consider such metrics in what follows.

The high causal fidelity values found for the SCGC networks suggests to consider
the topology as frozen during the time evolution of the system. The static
assumption allows us to use standard static centrality measures to rank the
users' influence.
Given a set of initial rumor spreaders (seeds), we are interested in quantifying
the dependence of those seeds out-degree $k_i^{out} = \sum_j A^T_{ij}$ on the
final outbreak size, varying $\beta$ and $\mu$.
The natural measure of the nodes outbreak size is the spreading capacity $q_i$
of node $i$, defined as the average number of nodes in the stifler state
(recovered) after the end of the infection process. This is defined as
\begin{equation}
    q_i \equiv \lim_{t \rightarrow \infty} \braket{Z^{(t)}}_i/N \, ,
\end{equation}
where the average $\braket{\dots}_i$ is evaluated over $10^2$ different
realization of the stochastic ISS dynamics described above, with the infection
originating at node $i$.

The initial condition is of one spreader located at node $i$ and all other nodes
ignorant. At each time step, the spreader nodes spread the rumor to their
ignorant neighbors with probability $\beta$ and turn into a stifler with
probability $\mu$ for each stifler neighbor. The dynamics stops when there are
no more spreaders in the network.

We choose $\beta=0.1$ and $\mu=1.0$ as the reference observation point in the
spreading parameter space, because is the closest at the decimal precision to
the critical epidemic threshold, assuming that it is vanishing.
The top 10 ranked users in terms of spreading capacity are presented in
Table~\ref{tab:influ_spreadcap} for the MN and RN, respectively.

\begin{figure}
    \centering
    {\includegraphics[scale=0.50]{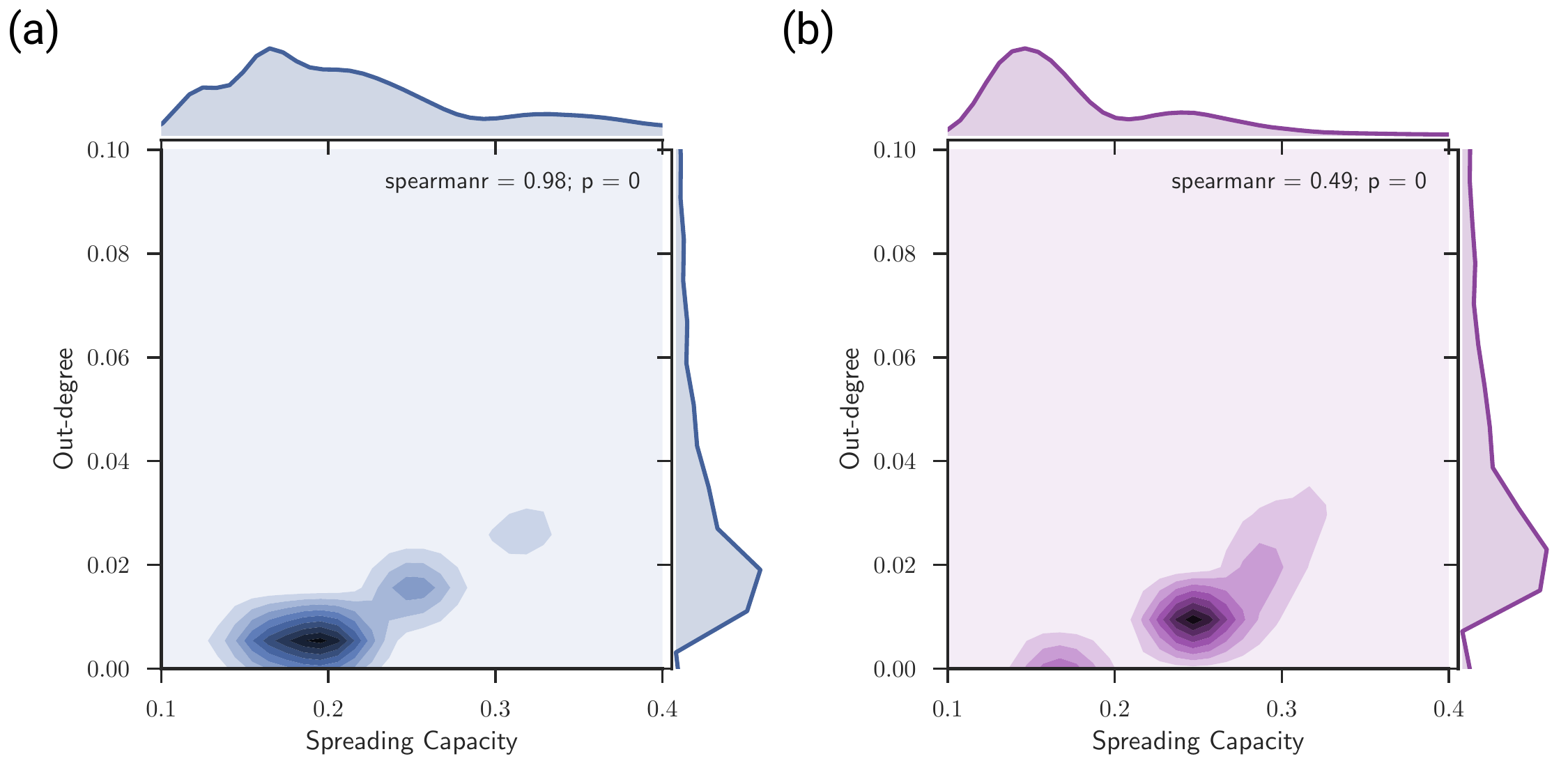}}
    \caption{
        \label{fig:figure8}
        Kernel density estimation of the correlation between the distributions
        of the max-normalized spreading capacity $q/\max[q]$ and the out-degree
        $k^{out}/\max[k^{out}]$ for the aggregated SCGC MN (blue) and RN
        (violet). Parameters values for the spreading and recovery rates are
        respectively  $\beta=0.1$ and $\mu=1.0$.
    }
\end{figure}

We found that the best performance metrics is the out-degree $k^{out}$, followed
by the closeness centrality and the k-core index $k_c$ for both MN and RN.
The Spearman's rank correlation coefficient between the spreading capacity and
the out-degree is shown in Figure~\ref{fig:figure8}.
The value of the correlation coefficient for the complete $\beta$ range with
unitary recovery is shown in Table~\ref{tab:spearMN}.

Besides the very bad performance of PageRank, the best metric to rank the
spreading ability of the users is the out-degree. Our results show a better
performance of all metrics apart from PageRank in the MN, while the gap between the
users' number of retweets and the other metrics is smoothed in the RN.
For the RN all metrics performed very poorly, which is probably due to the
strong segregation of the network compared to the MN. In the latter instead, a
local quantity such as the out-degree has proven to be sufficient to capture the
rumor dynamics to a high level of accuracy, as it displays an almost perfect
match with the spreading capacity ranking.

\begin{table}
    \centering
    \begin{tabular}{l|l|l}
        Rank & UserID (MN) &  UserID (RN)\\
        \hline
        \hline
        1 & @DartSirius & @Dani\_Gambit \\

        2 & @guffanti\_marco & @nuccioaltieri \\

        3 & @lorenzo3107 & @cadolo56 \\

        4 & @Alessandro02088 & @attanasio\_g \\

        5 & @mdpennalunga & @GiuliaPozzuoli \\

        6 & @alessandrab72 & @nonfraledonne \\

        7 & @onda\_di\_mare & @Bessico2 \\

        8 & @angeloargento & @rpp\_tweet \\

        9 & @fcerasani & @LaVarcaDiNoe \\

        10 & @pasqualegranata & @DCrognaletti \\
        \hline
    \end{tabular}
    \caption{
        \label{tab:influ_spreadcap}
        Top 10 ranked users for their spreading capacity for $\beta=0.1$ and
        $\mu=1.0$ for the MN (left) and the RN (right).
    }
\end{table}
\begin{table}
        \begin{tabular}{l||l|l|l|l|l|l||l|l|l|l|l|l}
            \multicolumn{1}{c||}{}&\multicolumn{6}{c||}{\textbf{Mentions Network (MN)}} & \multicolumn{6}{c}{\textbf{Retweets Network (RN)}}\\ 
            \hline
            $\beta$ & \textbf{$k^{out}$} & $b$ & $cl$ & $e$ & $k_c$ & $x$ & \textbf{$k^{out}$} & $b$ & $cl$ & $e$ & $k_c$ & $x$\\ 
            \hline
            $0.1$ & \textbf{0.98} & 0.63 & 0.83 & 0.36 & 0.83 & 0.15 & \textbf{0.49} & 0.38 & 0.48 & 0.20 & 0.45 & 0.22 \\
            $0.2$ & \textbf{0.97} & 0.63 & 0.83 & 0.35 & 0.83 & 0.15 & \textbf{0.48} & 0.37 & 0.46 & 0.20 & 0.44 & 0.22 \\
            $0.3$ & \textbf{0.97} & 0.63 & 0.83 & 0.35 & 0.82 & 0.15 & \textbf{0.48} & 0.37 & 0.46 & 0.19 & 0.44 & 0.21 \\
            $0.4$ & \textbf{0.96} & 0.62 & 0.82 & 0.35 & 0.82 & 0.15 & \textbf{0.48} & 0.37 & 0.46 & 0.19 & 0.44 & 0.22 \\
            $0.5$ & \textbf{0.96} & 0.62 & 0.82 & 0.36 & 0.82 & 0.15 & \textbf{0.48} & 0.37 & 0.46 & 0.19 & 0.44 & 0.21 \\
            $0.6$ & \textbf{0.95} & 0.62 & 0.81 & 0.35 & 0.81 & 0.15 & \textbf{0.48} & 0.36 & 0.46 & 0.18 & 0.43 & 0.21 \\
            $0.7$ & \textbf{0.94} & 0.61 & 0.81 & 0.35 & 0.80 & 0.15 & \textbf{0.49} & 0.37 & 0.46 & 0.20 & 0.44 & 0.22 \\
            $0.8$ & \textbf{0.93} & 0.61 & 0.80 & 0.35 & 0.80 & 0.15 & \textbf{0.49} & 0.37 & 0.46 & 0.19 & 0.43 & 0.22\\
            $0.9$ & \textbf{0.92} & 0.60 & 0.79 & 0.34 & 0.79 & 0.15 & \textbf{0.48} & 0.36 & 0.45 & 0.19 & 0.43 & 0.21\\
            $1.0$ & \textbf{0.92} & 0.60 & 0.79 & 0.34 & 0.78 & 0.14 & \textbf{0.48} & 0.36 & 0.45 & 0.19 & 0.43 & 0.21\\
            \hline
        \end{tabular}
    \caption{
        \label{tab:spearMN}
        Values of the Spearman's rank correlation coefficient  for the MN (left)
        and for the RN (right) in the full $\beta$ range at $\mu=1.0$ with the
        spreading capacity of the out-degree $k^{out}$, betweenness $b$,
        closeness $cl$, eigenvector $e$, k-core index $k_c$ and PageRank
        centrality $x$.
    }
\end{table}

\subsection*{Comparison with official polls}
\label{sec:comparisonPolls}

In the current section, we compare the daily temporal evolution of the average opinion
obtained from the Twitter dataset using the procedure described in
User dynamical opinion and the opinion trend obtained through the official polls.
The sources considered comprehend several sets of official polls (the whole
dataset is reported in the Supplementary Material). These
polls are carried out by different statistical research institutes and
commissioned by different customers such as: the official website of the
political and electoral polls of the Italian
Government, 
popular Italian newspapers such as La Stampa,
Il Corriere della Sera,
and Il Sole 24 ore,
and private monitoring companies.
The sample size of the surveys is heterogeneous but always statistically
significant, from a minimum of 400 to a maximum of 4000 individuals (about 1100
people interviewed on average).
In addition, the time span of the official polls we considered goes from the 31th of August
(the day we started collecting the data) to the 17th of November 2016, the day
before the pre-election silence started.

In our analysis, we compare the daily average opinion $\langle o_{t} \rangle =
N_{users}^{-1}\sum_{i=1}^{N_{users}} o_{i,t}$ with the official polls opinion,
defined as the difference between the proportion of pro-yes and pro-no voters in
each poll. In both the Twitter opinion and the official polls we considered also
the undecided users.
Note that the time span of the official polls is shorter than the Twitter one because the Italian law forces the official polls to stop two weeks before the vote. On the other hand, the Twitter data were recorded until the midnight of the 4th of December. Moreover, while the average opinion $\langle o_{t} \rangle$ measured from the tweets has a daily temporal resolution (as we project the daily user activity), the official polls do not have a regular frequency as they were
generally published every couple of days.

\begin{figure}
    \centering
    {\includegraphics[scale=0.30]{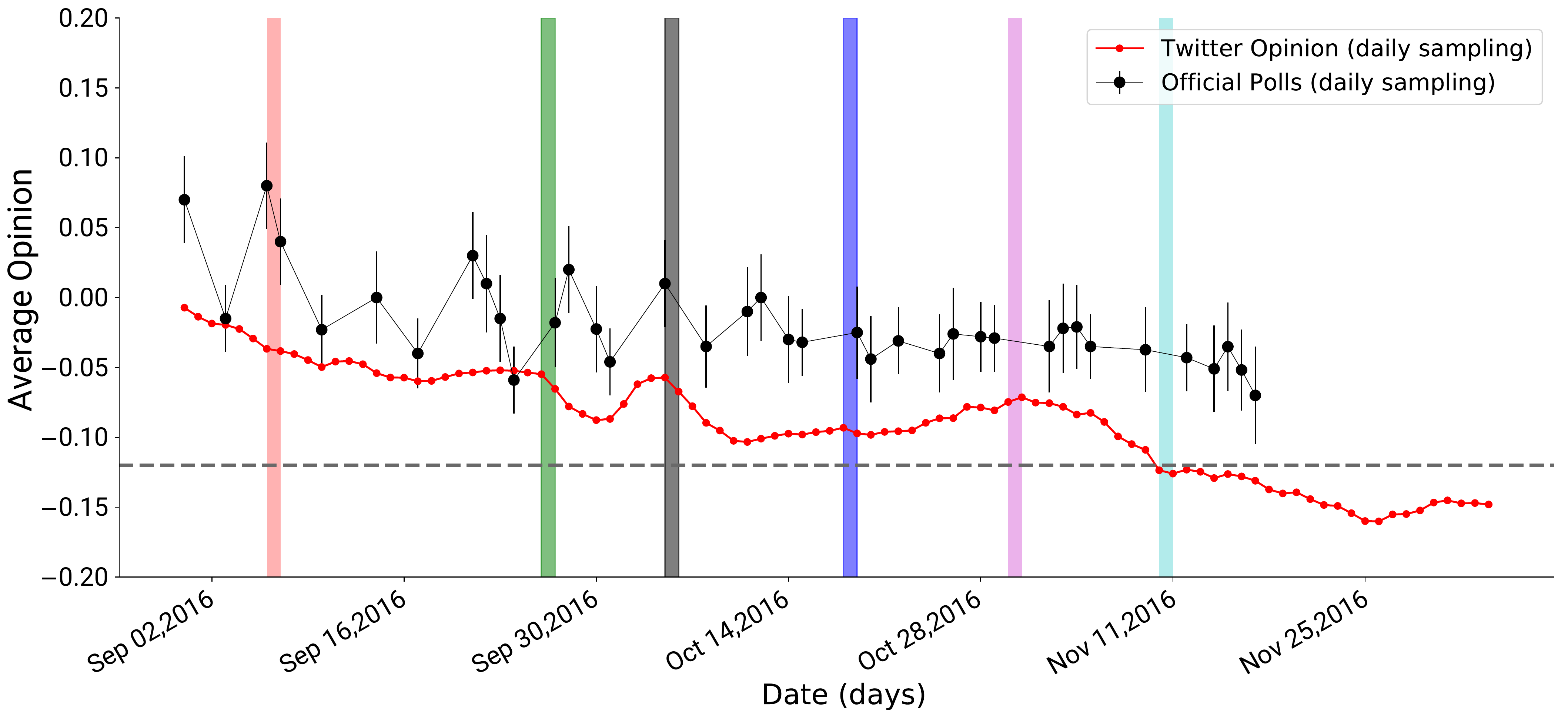}}
    \caption{
        \label{fig:figure9}
        The daily comparison between the variable $\sum_{i=1}^{N_{users}}
        o_{i,t}$ (red) and the opinion obtained by official polls (black). The
        error bars on the official polls data represents the statistical error
        range given in each poll. The black dashed line represents the final
        result of the voting day $-0.12$. The vertical lines represent some
        important events for the referendum debate, as described in the caption
        of \figurename~\ref{fig:figure3}.
    }
\end{figure}

\figurename~\ref{fig:figure9} shows the qualitative comparison, at the maximum
temporal resolution, between $\langle o_{t} \rangle$ and the opinion reported by
the official polls. In this plot, zero represent the perfect equilibrium between
the pro-yes and pro-no voters and a positive or negative value of the average
opinion represent respectively a majority of pro-yes or pro-no voters.
We observe that during the first sampling period, from August 31th to September 20th, the average opinion obtained through the official polls fluctuates around zero, reaching a maximum of $0.09$ when the major of Rome, who endorsed the No, was involved in legal issues (red bar in \figurename~\ref{fig:figure9}). On the other hand, the average opinion obtained through Twitter data starts near zero and then fundamentally decreases toward negative values.
After the 20th of September, when the Italian government fixed the official
voting day, the behaviour of the official polls starts to be more stable and
prone to a pro-no vote.
On the 5th of October, when the regional administrative court (\textit{tribunale
amministrativo regionale}, TAR) of region Lazio rejected a petition which had
requested a partial invalidation of the referendum, the trend changes in both
cases towards a more pro-yes political orientation. Subsequently, the two
opinions approach more negative values until the 17th of November, when the
official polls opinion is equal to $-0.07(\pm{0.03})$ while $\langle o_{t}
\rangle =-0.13$.

\begin{figure}
    \centering
    {\includegraphics[scale=0.50]{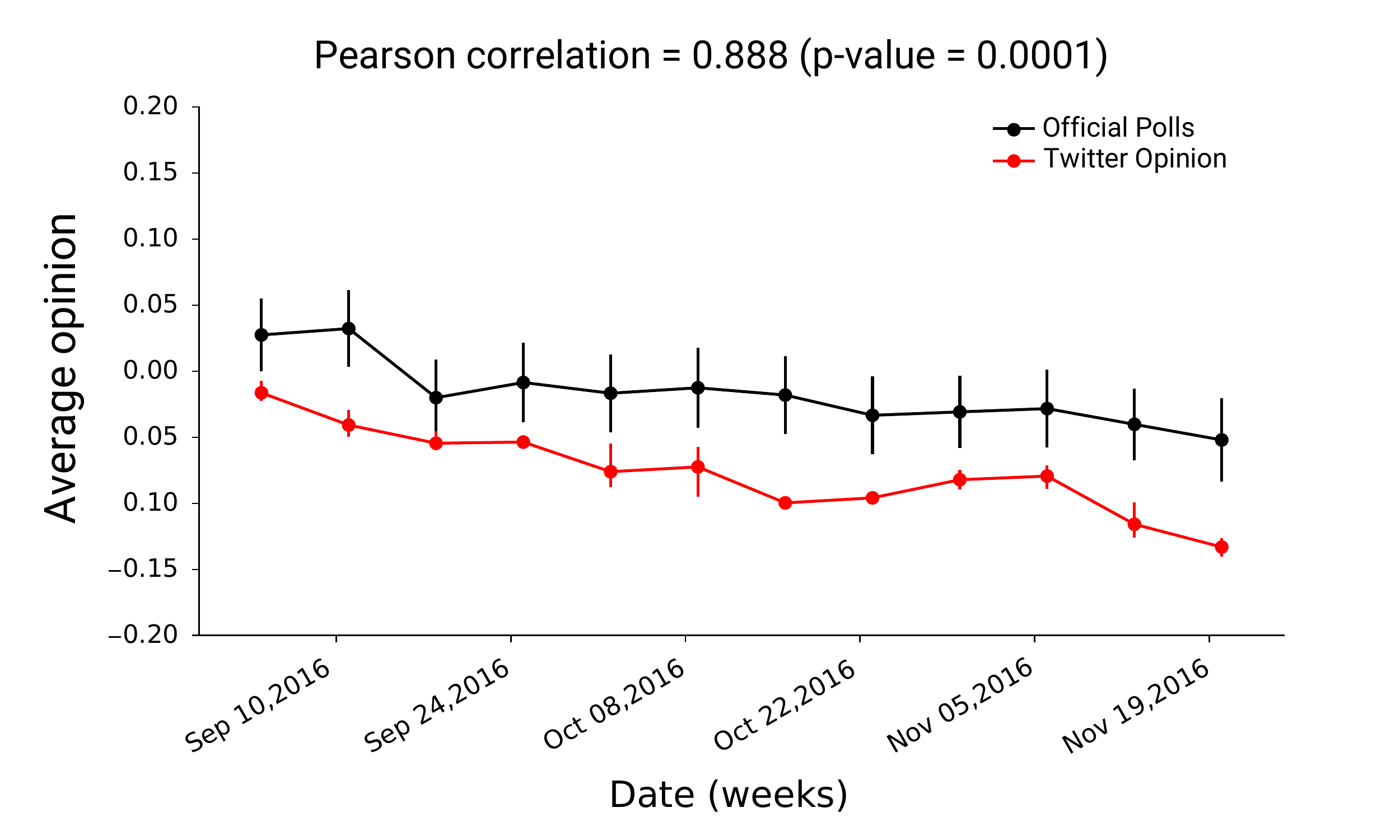}}
    \caption{
        \label{fig:figure10}
        Average opinion per week in the Official Polls (black) and in the
        Twitter Opinion (red) datasets. The error bars on the Twitter opinion
        (red) represent the minimum and the maximum value obtained in each week,
        while the error bars of the official polls (black) are obtained as the
        average error per week.
    }
\end{figure}

Finally, to compare the different trends in a quantitative way we re-sampled the
two time series averaging them with a weekly timescale. The resulting weakly
average opinion for both datasets is showed in \figurename~\ref{fig:figure10}. The
error bars on the Twitter opinion (red line) represent the minimum and the
maximum value obtained in each week, while the error bars of the official polls
(black line) are obtained as the average error per week. The Pearson
coefficient between the two time series is $r=0.888$ with a p-value equal to
$p=10^{-4}$, showing a high degree of correlation between the two time series.
Moreover, we find our latest reconstructed opinion value ($\langle o_{t} \rangle
= -0.15$) to outperform the official polls in predicting the final referendum
result. Indeed, given that $65.47\%$ of the eligible population voted and  $40.88\%$ of this percentage voted Yes and the remaining
$59.12\%$ voted No, the average opinion on the vote outcome is $-0.12$.

\section*{Conclusions}
\label{sec:conclusions}

In this work we measured and characterized the discussion about a political
event of national relevance in Italy using data from the  Twitter microblogging
platform.

We discussed the procedure implemented to collect tweets related to the Italian
constitutional referendum, which allowed us to obtain a large amount of data for
our analysis (approximately 7 millions tweets). Using a manually annotated
subset of tweets, we trained a classifier able to predict the leaning of tweets
with great accuracy ($86\%$ accuracy in a 4-fold cross validation). We deployed
this classifier to predict the opinion of each user in the system given the
user's history and activity. Notably, our definition is dynamical so that the
opinion of an user can change in time and it is not bound to a value computed at
the end of the observation period.

Thanks to the dynamical opinion, we performed a  characterization of the
interaction network topology in terms of the average opinion. We found strongly
polarized communities composed by users sharing the same opinion that internally
interact with retweets, and that interact with other communities only by
mentions.

The results are twofold: on the one hand they allow us to investigate the
influential spreaders and the relevant nodes of the retweet and mention
networks, while on the other hand they allow for a prediction of the population
opinion trend. The former point is here achieved using the temporal network of
interactions, thus without collecting the static network of friendships, an
operation that turns out to be unfeasible on such large networks.

It is worth noting that the influential spreaders identified by our method are
private users and not the official pro-yes and pro-no accounts as one would
expect. From the ranking correlation of the rumor-spreading capacity of the active users, our analysis shows a clear out-performance of the out-degree over all other measures, in particular in the mention network. Although we did not perform extensive study of other relevant centralities measures defined specifically for undirected networks, such as non-backtracking centrality and random-walk accessibility, among the various measures we also find the k-core and closeness centrality as relevant for the identification of influential spreaders in social networks. Differently from previous finding~\cite{de2014role}, in the social network of political discussion analyzed in this work, the simplest local measure of connectivity in the network, the out-degree $k^{out}$, is sufficient to estimate the correct ranking of users with extreme accuracy, and a correlation up to the value $r=0.98$, when approaching criticality.

Regarding the opinion trend, we found our estimate to be in good agreement with
official polls. Our method is particularly interesting as it is significantly
cheaper to track twitter activity rather than to finance a poll. Moreover,
Italian laws prohibits companies and parties to perform and publish polls in the
two weeks preceding a vote whereas, to the best of our knowledge, no restriction is currently given on Twitter
data. It is therefore possible to track and characterize the general opinion
dynamics in Twitter up to the date of an election, and for a longer time span
with respect to the official surveys.

Besides the opinion trend, our analysis also allows for the identification of
key-events influencing the overall opinion of the system. This can possibly
provide for a real-time investigation of the response of the population to some
public declarations or political events.

In conclusion, we developed a method to automatically collect and classify
politically relevant tweets and analyze their political leaning. Also, we are
able to evaluate the belonging of the users to certain discussion communities
and their reaction to relevant political events. We also showed that our method
gives a reliable prediction of the final outcome of the vote, serving this
purpose even better than the official polls in the event considered.





\section*{Competing interests}
The authors declare that no competing interests exist.

\section*{Author's contributions}
All authors conceived and designed the research project. All authors analyzed
the data and wrote the paper. All authors contributed equally to this work.

\section*{Funding}
This work has been founded by the DFG / FAPESP, within the scope of the IRTG 1740 / TRP 2015/50122-0.


\bibliography{ref}      

%


\input{SupplementaryMaterial.tex}

\end{document}

%% file: SupplementaryMaterial.tex
%

\newpage
\begin{center}
\textbf{\huge Supplementary Material}
\end{center}
\setcounter{equation}{0}
\setcounter{figure}{0}
\setcounter{table}{0}
\setcounter{page}{1}
\setcounter{section}{0}
\makeatletter
\renewcommand{\theequation}{S\arabic{equation}}
\renewcommand{\thefigure}{S\arabic{figure}}
\renewcommand{\thepage}{s\arabic{page}}

\section{Defining $K$ and $\epsilon$}
\subsubsection*{Qualitative analysis}
\begin{figure}
    \centering
        {\includegraphics[scale=0.3]{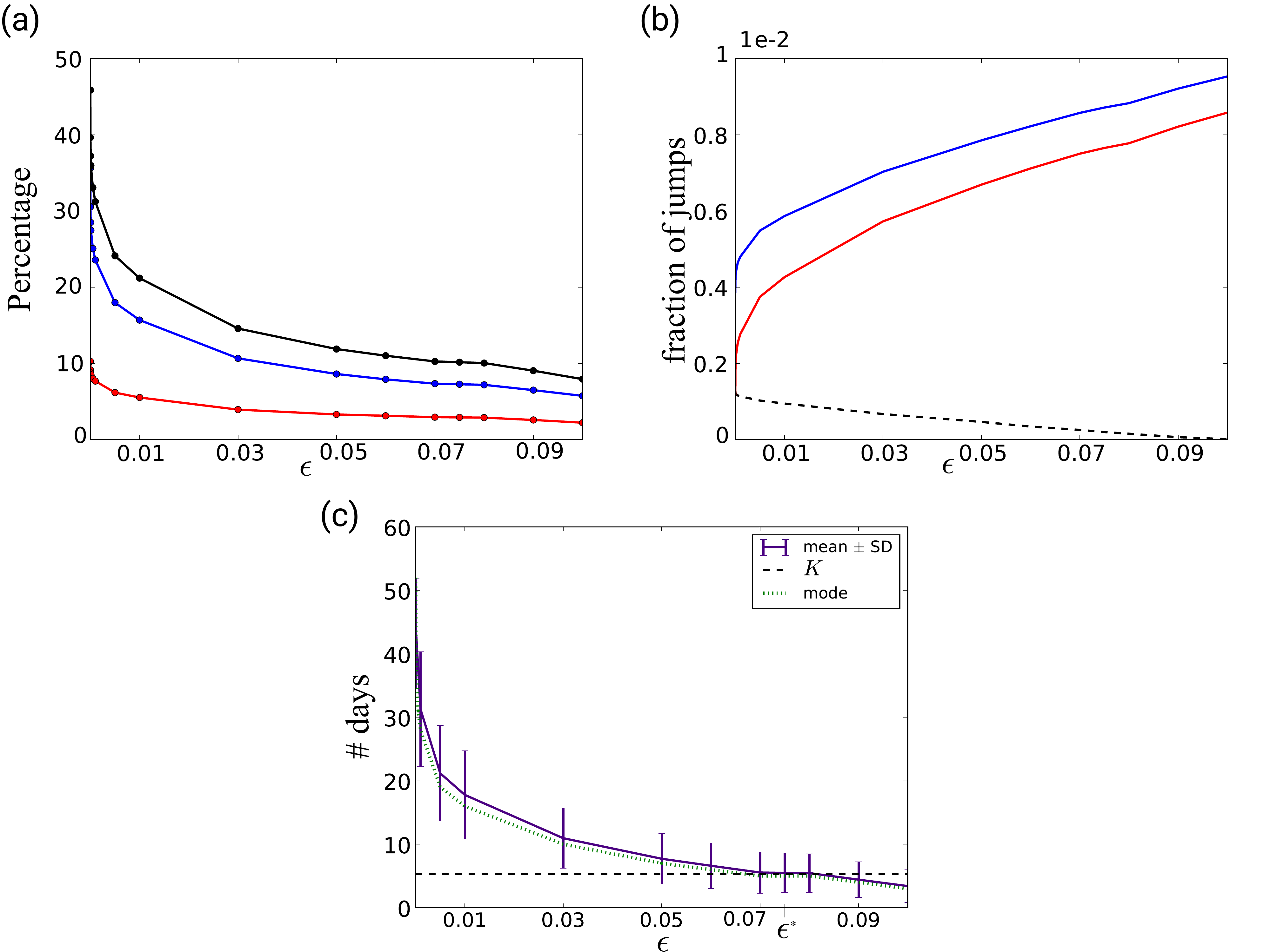}}
    \caption{
        \label{DefOp:fig:propertiesVSepsilon}
        Global properties of the opinion dynamics defined in (2)
        as a function of $\epsilon$.
        (a) In red (blue), the percentage of users that forever maintain a $+1$
        ($-1$) opinion after they first support Yes (No). In black, the sum of
        the two percentages.
        (b) The percentage of abrupt jumps (dashed black line) plotted against
        the percentage of jumps that approach Yes (red plain line) and the
        percentage of jumps that approach No (blue plain line).
        (c) Statistics of the number of days an opinion is maintained once it is
        assumed. The purple line shows the mean of such memory length and the
        associated vertical purple lines show the amplitude of the standard
        deviation. The dotted green line shows the mode, i.e. the value
        occurring most frequently. The horizontal black dashed line denotes the
        value of parameter $K= 5.29$ days inferred from the dataset.
    }
\end{figure}

To choose a proper value of $\epsilon$, we tested the outcome of the global
process considering different values of $\epsilon$ in the interval $[0,0.1]$.
First, we consider the percentage of users that forever maintain a polarized
opinion after they assume one, who we name stubborn users. As hypothesized (see main text),
\figurename~\ref{DefOp:fig:propertiesVSepsilon} panel a shows that the overall
percentage halves when $\epsilon$ increases from zero to $0.01$, and it further
reduces for larger values of the parameter. Another predictable effect of
increasing $\epsilon$ is the reduction of the number of abrupt opinion jumps
(i.e. direct switch from supporting Yes to supporting No, and \textit{vice}
\textit{versa}).
This is illustrated in \figurename~\ref{DefOp:fig:propertiesVSepsilon} panel b, which shows that the percentage of abrupt opinion jumps over the total amount of possible
jumps slowly decreases as a function of $\epsilon$ (black dashed line), passing
from about $0.1\%$ for $\epsilon=0$ to essentially zero percent for
$\epsilon=0.1$. This is a desirable property as it seems rather unrealistic that
a given user changes her/his mind all of a sudden, without passing through an
intermediate neutral state before.

However, panel b in \figurename~\ref{DefOp:fig:propertiesVSepsilon}  also reveals that the
increase of both the number of opinion jumps towards a Yes vote (i.e. opinion
jumps from No to Neutral and from Neutral to Yes, in red) and towards a No vote
(from Yes to Neutral and from Neutral to No, in blue) largely counterbalance the
reduction of abrupt jumps, yielding to a net increment in total number of
jumps.
As a consequence, the overall dynamics is less stable, resulting in users who
change opinion more often for large values of $\epsilon$. In fact, also the
maximum number of user's opinion jumps among all users increases as a function
of $\epsilon$, passing from twenty-five jumps for $\epsilon=0$ to a maximum
number of over forty-five opinion jumps for $\epsilon=0.1$ (not shown in the
picture). Although a richer and more complex dynamics might seem more
interesting, it is rather unlikely that a person changes is mind this
frequently. We draw the conclusion that $\epsilon$ should not be too large in
order to reduce users' flicking.

To summarize, in this section we demonstrated that the effect of increasing the parameter
$\epsilon$ is, on the one hand, to reduce the users' stubbornness (the tendency to
maintain a given opinion once it is assumed). On the other hand, a larger value
of $\epsilon$ results in a smaller number of abrupt jumps between Yes and No, at
the cost of larger users' opinion variability. However, we still lack a quantitative method allowing to asses the value of $\epsilon$ that produces a good trade-off between users' stubbornness and opinion volatility.

\subsubsection*{Quantitative analysis}
To this aim, we observe that $\epsilon$ also has an effect on the memory of
users' past tweets. Indeed, since stubbornness and opinion retention are related,
panel a in \figurename~\ref{DefOp:fig:propertiesVSepsilon} suggests that smaller
values of $\epsilon$ result in longer users' memory.
In order to quantify such memory length, we considered the mean and the mode of
the number of days an opinion is preserved once it is assumed. Such statistics
are computed solely considering those users that acquired an opinion only once
in the dataset, and then lost it.
We decided not to consider users that assume and lose an opinion multiple times
in order to focus on the effect of small clusters of Tweets which are frequent
enough not to let the opinion vanish. Also, the requirement that the opinion has
to be abandoned eventually rules out users that are ideologically polarized and
would significantly extend the memory length  statistics (an instance of ideologically polarized users are the official accounts represented in 
\figurename~2 
panels a and b).

In conclusion, \figurename~\ref{DefOp:fig:propertiesVSepsilon} panel c shows that only for
$0.07<\epsilon<0.08$ the memory length is reasonably close the value of $K=5.29$
days which we identify in the main text. Hence, we analyzed the
opinion dynamics of some users who only tweeted sporadically in favour of either
Yes or No, and such analysis corroborated the choice of $\epsilon^{*}=0.075$.

\begin{figure}
    \centering
            {\includegraphics[scale=0.5]{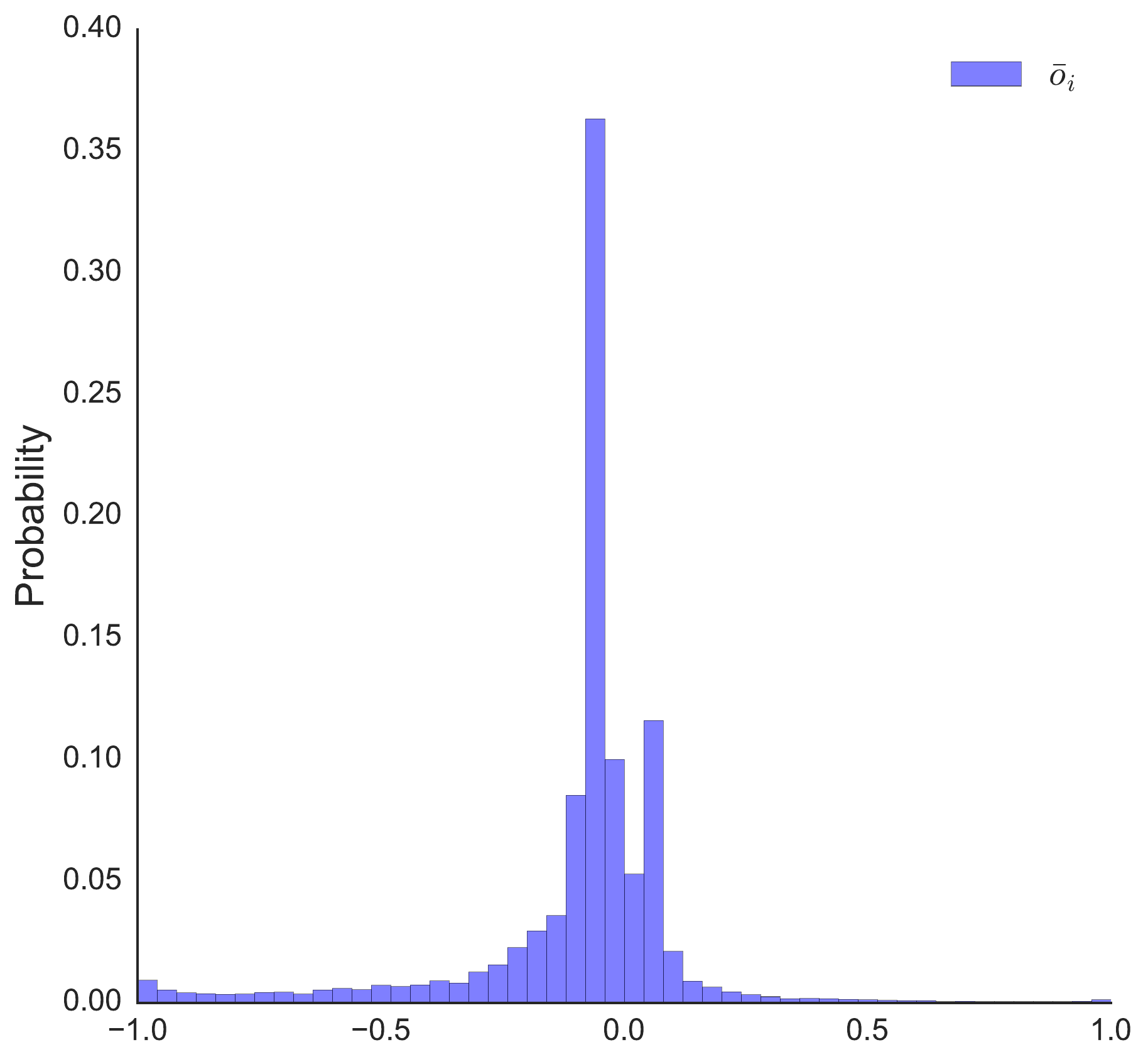}}
    \caption{
        \label{fig:hist}
        The distribution of the time average opinion of the users $\bar{o}_i$.
    }
\end{figure}

As a synthetic description of the opinion resulting from the choice of $K=5.29$ days and $\epsilon=0.075$,  in \figurename~\ref{fig:hist} we show the empirical distribution (histogram) of the users' time-averaged opinion 
$$\bar{o}_i = T^{-1}\sum_{t=1}^T o_{i,t}, \qquad i=1,\dots,N_{users}$$
which exhibits an evident skewness towards negative values (\textit{i.e.} towards pro-no opinions).

\section{Detailed description of the histograms in \figurename~2}  

In \figurename~2, 
the $y$-axis represents the time
spanning from $t=1$ (31st August 2016) to $t=97$ (the date of the referendum),
whereas the $z$-axis counts the number of tweets authored by the user on a
specific day. Such histograms are subdivided according to how the tweets were
classified. In fact,  the $x$-axis contains the possible tweets classifications:
$-1$ for a tweet supporting No (in red), $0$ for a neutral or irrelevant tweet
(gray), and $+1$ for a tweet supporting Yes (blue).

\figurename~2 
panel a shows the histogram for the daily
activity of the official pro-yes committee's account @bastaunsi (twitter
id 733695386846662657), whereas \figurename~2 
panel b represents one of the official pro-no committees' account @comitatono
(twitter id 696674734969397248). The line lying on the right side of the
$xy$-plane shows the resulting user's opinion time course, where blue stays for
a pro-yes opinion, gray for a neutral opinion, and red for a
pro-no opinion.

For both accounts we can observe a continuous twitting activity supporting the
relative pole. As expected, such a sustained activity results in an opinion
which is neutral at the beginning (neither of the accounts twitted on August
31st) but it changes for good once the first polarized tweet is posted. It is
interesting to remark that the account @bastaunsi produced a much larger number
of tweets than @comitatono, with a maximum of over one hundred tweets per days for the official pro-yes account versus a maximum of four tweets per day for the official pro-no account. 
Moreover, note that although the machine-learning algorithm
classified some of @bastaunsi tweets as pro-no (note the red bars on the left of
\figurename~2 
panel a), the large mole of pro-yes
tweets allows to overlook what is likely to be a misclassification of the
machine-learning algorithm.

Finally, \figurename~2 
panel c shows the daily activity histograms for account @cechidiceno27 (twitter id 780796377148162048). 
We observe that such user only started posting at $t=40$ (9th October 2016) with a pro-yes tweet, which resulted in a pro-yes opinion that lasted three days. Then, because of the repeated pro-no activity, its opinion returns to be neutral for a couple of days, just to settle definitively on a $-1$ opinion at $t=45$.

\section{Analysis of the temporal network for the rumor spreading dynamics and network centrality}

\begin{figure}
    \centering
            {\includegraphics[scale=0.4]{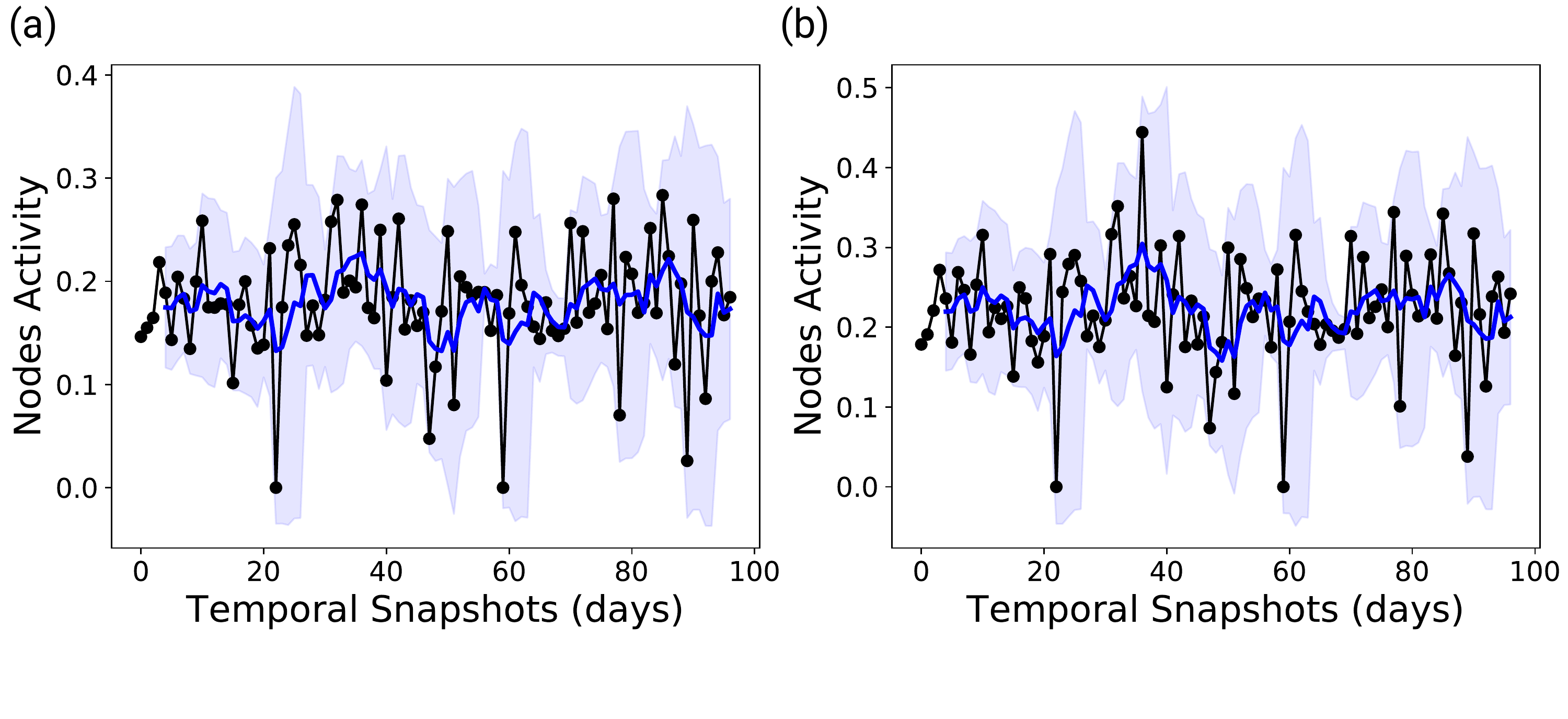}}
    \caption{
        \label{fig:activity}
        Fraction of active nodes  in the temporal observation window of $T=97$
        days for the SCGC MN (a) and SCGC RN (b). Two days with zero net
        activity were days of missed data collection.
    }
\end{figure}

The temporal nature of the dataset presents many challenges, first and foremost
\emph{causality}. To spread information from node $j$ to node $k$ at time
$t'=t+dt$, the same information must have already reached node $j$ from node $i$
at a previous time step $t<t'$. Therefore the time-aggregation of the temporal
networks can include paths of rumor propagation that are not present in the
causally ordered temporal sequence of contacts.
As it can be seen in Figure~\ref{fig:activity} panels a and b, the
temporal activity of the SCGC (\textit{i.e.} the fraction of nodes with at least one
active edge at each time) is fluctuating during the whole time window.

To quantify the impact of causality driven by the temporal feature of the
network, following \cite{lentz2013unfolding}, we compute the causal fidelity
$c\in[0,1]$ of the SCGC graphs. The causal fidelity is defined as the fraction
of the number of paths in the time-aggregated static network which can be also
taken in the temporal one. Thus
$c=\rho(\mathbf{\mathcal{A}}^T)/\rho(\mathbf{A}^T)$ where
$\mathbf{\mathcal{A}}^T = \bigwedge_t^T (\mathbf{I} + \mathbf{A}^{(t)})$ and
$\mathbf{A}^T = \bigvee_t^T \mathbf{A}^{(t)}$ are the full aggregated and
accessibility boolean matrices obtained with the boolean product and sum
respectively, and $\rho(\mathbf{A}^{(t)}) = \sum_{ij} A^{(t)}_{ij}/N^2$ is the
density of the $t$-th snapshot. The maximal causal fidelity $c=1$ implies that
the temporal and static representations share the same path density. On the
contrary, a low value of $c$ indicates that most of the paths in the static
aggregate approximation do not follow a causal sequence of edges and thus do not
belong to the temporal network.
A fundamental quantity that we use in the main text is
$\rho(\mathbf{A}^{(t)})-\rho(\mathbf{A}^{(t-1)})$ that gives the probability to
find a path of length $t$, with $t=1,\dots,T$, between two randomly chosen
nodes.

\begin{figure}
    \centering
            {\includegraphics[scale=0.5]{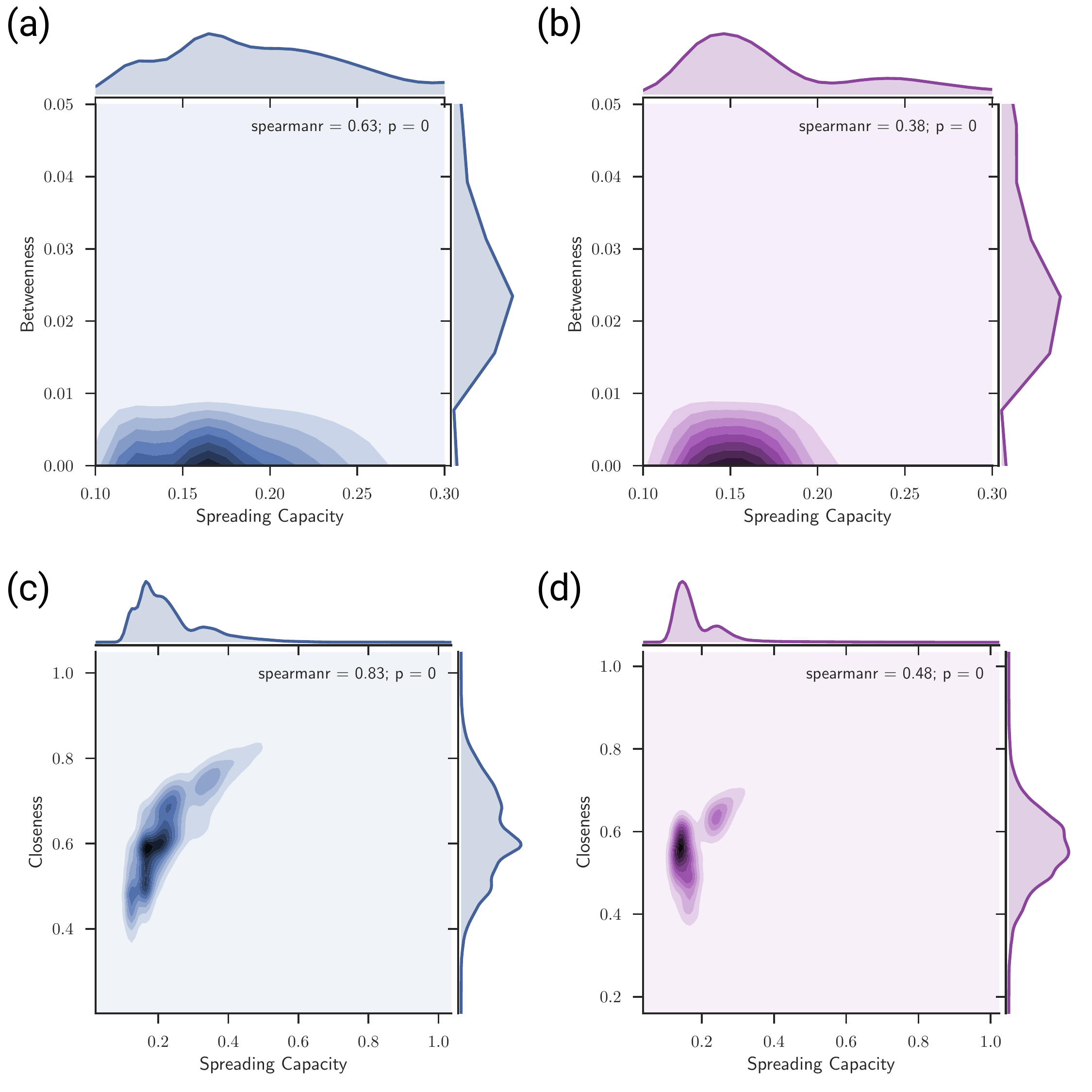}}
    \caption{
        \label{fig:corra}
        Kernel density estimation of the correlation between the distributions
        of the max-normalized spreading capacity $q/\max[q]$ and the
        max-normalized betweenness (top) and closeness (bottom) for the
        aggregated SCGC MN (blue) and RN (violet). Parameters values for the
        spreading and recovery rates are respectively  $\beta=0.1$ and
        $\mu=1.0$.
    }
\end{figure}

\begin{figure}
    \centering
            {\includegraphics[scale=0.5]{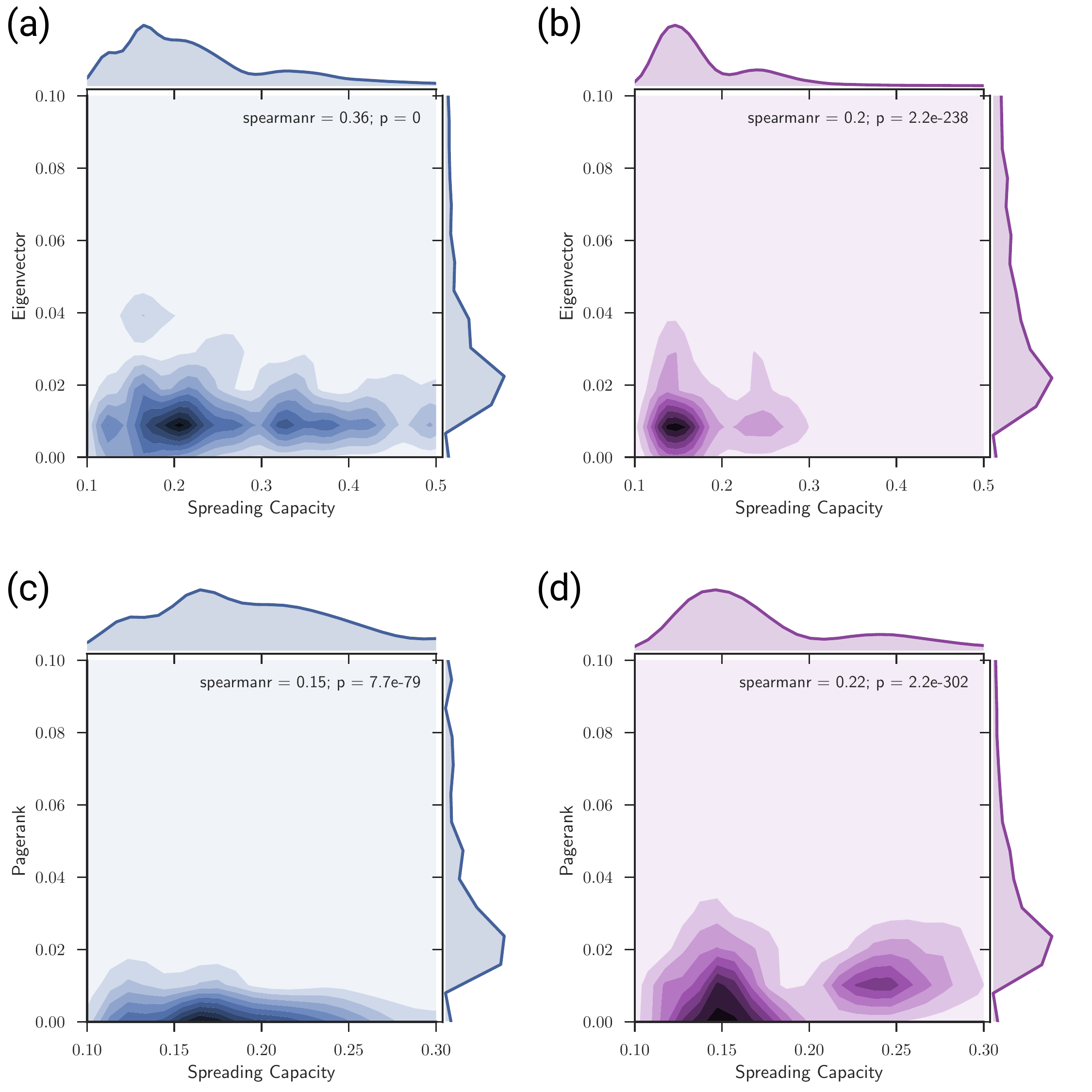}}
    \caption{
        \label{fig:corrb}
        Kernel density estimation of the correlation between the distributions
        of the max-normalized spreading capacity $q/\max[q]$ and the
        max-normalized eigenvector (top) and PageRank (bottom) for the
        aggregated SCGC MN (blue) and RN (violet). Parameters values for the
        spreading and recovery rates are respectively  $\beta=0.1$ and
        $\mu=1.0$.
    }
\end{figure}

\begin{figure}
    \centering
            {\includegraphics[scale=0.5]{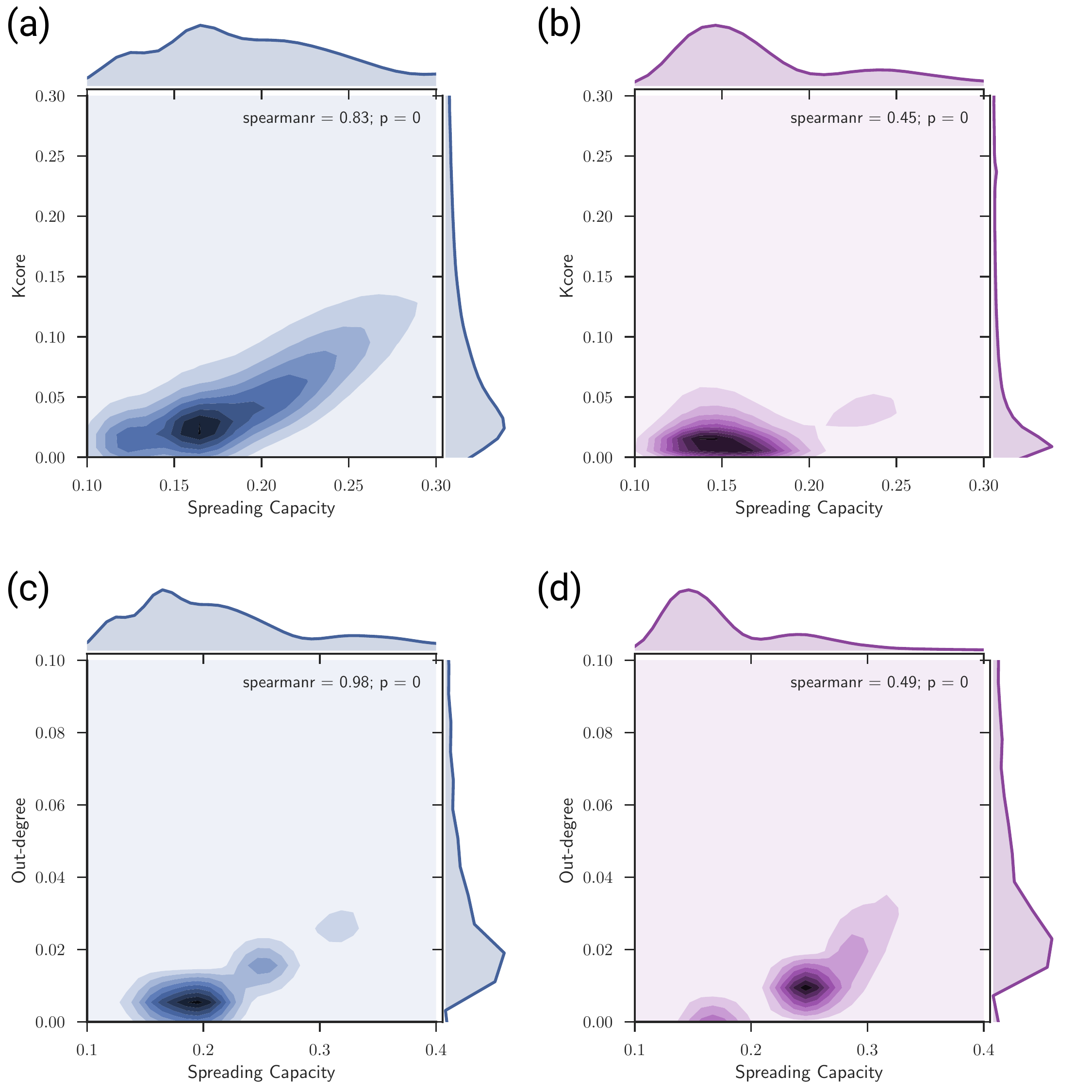}}
    \caption{
        \label{fig:corrc}
        Kernel density estimation of the correlation between the distributions
        of the max-normalized spreading capacity $q/\max[q]$ and the
        max-normalized k-core (top) and out-degree (bottom) for the aggregated
        SCGC MN (blue) and RN (violet). Parameters values for the spreading and
        recovery rates are respectively  $\beta=0.1$ and $\mu=1.0$.
    }
\end{figure}

We now briefly review the definition of the centrality measures used in the main text. For a detailed description see \cite{newman2010networks}, for example.

The degree centrality $k$ is defined as the number of connections of each vertex. For a directed network we have the distinction between the in-degree and out-degree. The latter, which is relevant for our discussion, is the sum of each row  of the adjacency matrix $A_{ij}$
\begin{equation}
    k^{out}_i = \sum_j A_{ij}.
\end{equation}
The betweenness centrality $b$ is a measure of the amount of information transmitted by shortest paths along each node. It is defined as the fraction of shortest paths between all pairs of nodes that pass through node $i$
\begin{equation}
    b_i = \sum_{kl} g_{kl}(i)/g_{kl},
\end{equation}
where $g_{kl}$ is the total number of shortest paths from node $k$ to node $l$ and  $g_{kl}(i)$ is the number
of these paths that also pass through node $i$.

The closeness centrality $cc$ is the reciprocal of the  mean shortest-path length $l_{ij}$ from a node to all others 
\begin{equation}
    cc_i = N/\sum_j l_{ij}.
\end{equation}
The eigenvector centrality $e$ is
defined by the components of the leading eigenvector of the adjacency matrix 
\begin{equation}
    e_i = \lambda_1^{-1}\sum_{j}A_{ij} e_j,
\end{equation}
where $\lambda_1$ is the leading eigenvalue of $\mathbf{A}$. 

The k-core decomposition of a graph is obtained from the subgraphs composed of nodes that have at least $k$ neighbors within the set itself for all possible values of the nodes degree $k$. For directed graphs one assumes that $k=k^{in} + k^{out}$. The k-core centrality $k_c$ of a node then equals the largest value of k-cores which the node belongs to.

Similarly to the eigenvector centrality the PageRank centrality $x$ is defined in terms of the
spectra of the so-called Google matrix. PageRank can be computed iteratively from  
\begin{equation}
    x_i = \alpha\sum_{j}A_{ji} x_j /k_j^{out} + (1-\alpha)/N,
\end{equation}
where $\alpha=0.85$ is the teleportation parameter, which gives the probability to randomly relocate during a random-walk in the network.

In Figure~\ref{fig:corra}, ~\ref{fig:corrb} ~\ref{fig:corrc} we show the kernel
density estimation of the spreading capacity with the centrality measures.


\begin{figure}
    \centering
            {\includegraphics[scale=0.1]{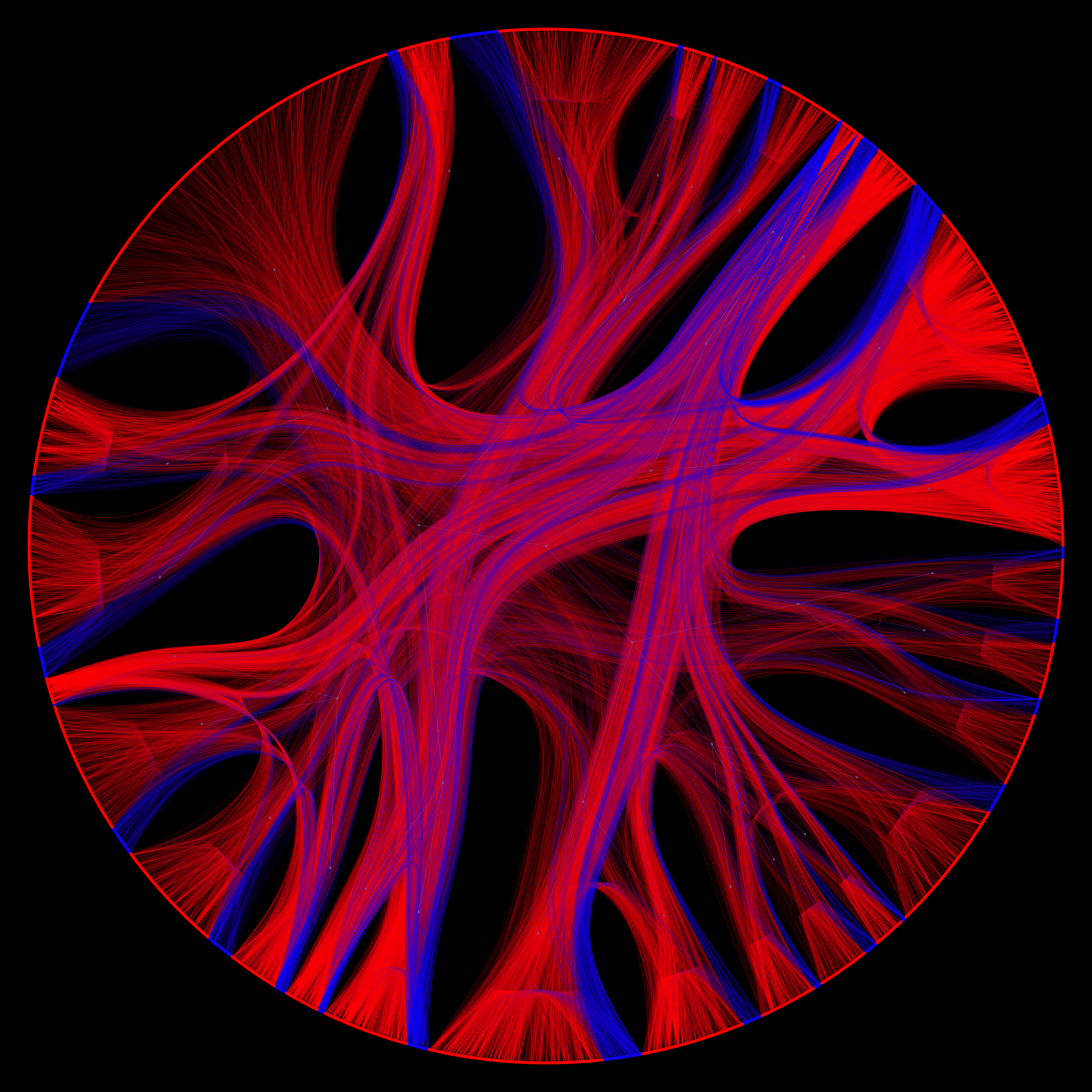}}
    \caption{
        \label{fig:opinion1}
        Circular representation of the strongly connected component of the
        aggregated network of mentions with the time average of the users'
        opinions $\bar{o}_i$. The average opinion is represented here with color
        codes as blue (Yes) if $\bar{o}_i>0$  and red (No) if $\bar{o}_i\le
        0$, and analogously for the edges. The node's ordering is given by the
        stochastic block model for community detection
        \cite{decelle2011asymptotic,peixoto_graph-tool_2014,peixoto2014hierarchical}.
    }
\end{figure}

\section{Official polls dataset}\label{section:Dataset}

The official polls dataset is available in the file \texttt{ReferendumOfficialPolls.csv}.

\clearpage

\bibliography{ref}